\documentclass[twocolumn,twocolappendix]{aastex631}
\graphicspath{{./fig/}{./png/}}

\usepackage{amsmath}

\newcommand{\EQ}{\begin{equation}}
\newcommand{\EN}{\end{equation}}
\newcommand{\EQA}{\begin{eqnarray}}
\newcommand{\ENA}{\end{eqnarray}}

\newcommand{\EEq}[1]{Equation~(\ref{#1})}
\newcommand{\Eq}[1]{Equation~(\ref{#1})}

\newcommand{\Sec}[1]{Section~\ref{#1}}

\newcommand{\Fig}[1]{Figure~\ref{#1}}

\newcommand{\Figp}[2]{Figure~\ref{#1}({#2})}

\newcommand{\Tab}[1]{Table~\ref{#1}}

\newcommand{\App}[1]{Appendix~\ref{#1}}

\newcommand{\bra}[1]{\langle #1\rangle}

{}
{}

\newcommand{\tildeee}{\tilde{\mathbf{e}}}
\newcommand{\tildeAA}{\tilde{\mathbf{A}}}

\newcommand{\tildeA}{\tilde{A}}

\newcommand{\Iota}{\gamma}
\newcommand{\kk}{\mathbf{k}}

\newcommand{\xx}{\mathbf{x}}

\newcommand{\BB}{\mathbf{B}}

\newcommand{\EE}{\mathbf{E}}

\newcommand{\AAA}{\mathbf{A}}

\newcommand{\uu}{\mathbf{u}}

\newcommand{\nab}{{\mathbf{\nabla}}}

\newcommand{\TTT}{{\sf T}}
\newcommand{\tildeh}{\tilde{h}}
\newcommand{\tildeT}{\tilde{T}}

\newcommand{\tildeTTT}{\tilde{\sf T}}
\newcommand{\ii}{{\rm i}}

\newcommand{\dd}{{\rm d} {}}

\newcommand{\const}{{\rm const}  {}}

\def\la{\mathrel{\mathchoice {\vcenter{\offinterlineskip\halign{\hfil
$\displaystyle##$\hfil\cr<\cr\sim\cr}}}
{\vcenter{\offinterlineskip\halign{\hfil$\textstyle##$\hfil\cr<\cr\sim\cr}}}
{\vcenter{\offinterlineskip\halign{\hfil$\scriptstyle##$\hfil\cr<\cr\sim\cr}}}
{\vcenter{\offinterlineskip\halign{\hfil$\scriptscriptstyle##$\hfil\cr<\cr\sim\cr}}}}}

\def\aa{\alpha}
\def\H{\mbox{\rm H}}

\def\Imag{\mbox{\rm Im}}

\def\EEK{{\cal E}_{\rm K}}
\def\EEM{{\cal E}_{\rm M}}
\def\EEEM{{\cal E}_{\rm EM}}
\def\EEEl{{\cal E}_{\rm E}}

\def\EEGW{{\cal E}_{\rm GW}}
\def\OmGW{{\Omega}_{\rm GW}}

\def\EGW{E_{\rm GW}}

\def\EK{E_{\rm K}}
\def\EM{E_{\rm M}}
\def\EEl{E_{\rm E}}

\def\qM{q_{\rm M}}
\def\qEM{q_{\rm EM}}

\def\xiM{\xi_{\rm M}}

\def\hrms{h_{\rm rms}}

\def\hc{h_{\rm c}}
\def\Tr{T_{\rm r}}
\def\kpeak{k_{\rm *}}

\def\HM{H_{\rm M}}
\def\EM{E_{\rm M}}

\def\kB{k_{\rm B}}

\def\kB{k_{\rm B}}

\def\Brms{B_{\rm rms}}

\newcommand{\W}{\,{\rm W}}
\newcommand{\V}{\,{\rm V}}

\newcommand{\T}{\,{\rm T}}
\newcommand{\G}{\,{\rm G}}

\newcommand{\Hz}{\,{\rm Hz}}
\newcommand{\mHz}{\,{\rm mHz}}

\newcommand{\nHz}{\,{\rm nHz}}

\newcommand{\K}{\,{\rm K}}
\newcommand{\g}{\,{\rm g}}
\newcommand{\s}{\,{\rm s}}

\newcommand{\GeV}{\,{\rm GeV}}
\newcommand{\MeV}{\,{\rm MeV}}

\newcommand{\m}{\,{\rm m}}

\newcommand{\J}{\,{\rm J}}

\newcommand{\A}{\,{\rm A}}

\begin{document}

\title{Simulations of helical inflationary magnetogenesis and gravitational waves}

\correspondingauthor{Axel Brandenburg}
\email{brandenb@nordita.org}
\author[0000-0002-7304-021X]{Axel Brandenburg}
\affiliation{Nordita, KTH Royal Institute of Technology and Stockholm University,
Hannes Alfv\'ens v\"ag 12, SE-10691 Stockholm, Sweden}
\affiliation{Department of Astronomy, AlbaNova University Center,
Stockholm University, SE-10691 Stockholm, Sweden}
\affiliation{McWilliams Center for Cosmology \& Department of Physics,
Carnegie Mellon University, Pittsburgh, PA 15213, USA}
\affiliation{School of Natural Sciences and Medicine, Ilia State University,
3-5 Cholokashvili Avenue, 0194 Tbilisi, Georgia}

\author[0000-0001-6082-0615]{Yutong He}
\affiliation{Nordita, KTH Royal Institute of Technology and Stockholm University,
Hannes Alfv\'ens v\"ag 12, SE-10691 Stockholm, Sweden}
\affiliation{Department of Astronomy, AlbaNova University Center,
Stockholm University, SE-10691 Stockholm, Sweden}

\author[0000-0002-2549-6861]{Ramkishor Sharma}
\affiliation{Inter University Centre for Astronomy and Astrophysics,
Post Bag 4, Pune University Campus, Ganeshkhind, Pune 411 007, India}

\begin{abstract}
Using numerical simulations of helical inflationary magnetogenesis in
a low reheating temperature scenario, we show that the magnetic energy
spectrum is strongly peaked at a particular wavenumber that depends
on the reheating temperature.
Gravitational waves (GWs) are produced at frequencies between $3\nHz$
and $50\mHz$ for reheating temperatures between $150\MeV$ and
$3\times10^5\GeV$, respectively.
At and below the peak frequency, the stress spectrum is always found to
be that of white noise.
This implies a linear increase of GW energy per logarithmic wavenumber
interval, instead of a cubic one, as previously thought.
Both in the helical and nonhelical cases, the GW spectrum is followed
by a sharp drop for frequencies above the respective peak frequency.
In this magnetogenesis scenario, the presence of a helical term extends
the peak of the GW spectrum and therefore also the position of the
aforementioned drop toward larger frequencies compared to the case
without helicity.
This might make a difference in it being detectable with space
interferometers.
The efficiency of GW production is found to be almost the same as in
the nonhelical case, and independent of the reheating temperature,
provided the electromagnetic energy at the end of reheating is fixed to
be a certain fraction of the radiation energy density.
Also, contrary to the case without helicity, the electric energy is now
less than the magnetic energy during reheating.
The fractional circular polarization is found to be nearly hundred per
cent in a certain range below the peak frequency range.
\end{abstract}

\keywords{gravitational waves---early Universe---turbulence---magnetic fields---MHD}

\section{Introduction}

There has been significant interest in the production of helical
magnetic fields and circularly polarized gravitational waves (GWs)
from the early Universe \citep{carroll1992, cornwall1997, vachaspati2001,
2005PhRvL..95o1301K, 2021PhRvR...3a3193K, Anber+Sorbo06, campanelli2008,
durrer2011, Caprini+Sorbo14, adshead2016, adshead2018}.
Owing to magnetic helicity conservation, such fields would have had a better
chance to survive until the present time \citep{axel2001, banerjee2004,
kahniashvili2016, axel2017}.
The associated electromagnetic (EM) stress also drives circularly
polarized GWs \citep{2005PhRvL..95o1301K, 2021PhRvR...3a3193K, Ellis+20,
RoperPol+21}.
If the sign and spectral shape of the circular polarization can in
future be detected, it would provide important information about the
underlying mechanisms responsible for the generation.

Inflationary magnetogenesis scenarios are particularly attractive,
because they have the advantage of producing large-scale magnetic fields.
They tend to amplify magnetic fields from quantum fluctuations by
the breaking of conformal invariance through a function $f$ such
that the Lagrangian density has a term that takes the form
$f^2 F_{\mu\nu} F^{\mu\nu}$, where $F_{\mu\nu}$ is the Faraday
tensor \citep{1988PhRvD..37.2743T,1992ApJ...391L...1R}.
However, those mechanisms can only be viable if they avoid
some well-known problems discussed in detail in the literature
\citep{mukhanov2009, Ferreira+13, Kobayashi+Afshordi14,
Kobayashi+Sloth19}.
These problems are avoided by requiring the function $f$ to
obey certain constraints that have been discussed in detail by
\cite{Sharma+17}. 
For some scenarios, these magnetic fields can lead to the production of
GWs which lie in the sensitivity range of space interferometers such as
LISA and Taiji, as studied analytically in \cite{2020PhRvD.101j3526S}.
This magnetogenesis model was then extended to the helical case
\citep[][hereafter referred to as SSS]{Sharma+18}.
A similar model of helical magnetogenesis was also considered by
\cite{fujita2019} and \cite{OF21}.
Numerical simulations have recently been performed for the nonhelical
case \citep[][hereafter BS]{Bran+Shar21}.
The goal of the present paper is to apply numerical simulations now to
helical magnetogenesis.
These models continue to amplify EM fields during the
post-inflationary matter-dominated era after inflation, but require
relatively low reheating temperatures, $T_{\rm r}$.
Values of $T_{\rm r}$ in the range of the electroweak and quantum
chromodynamics (QCD) epochs are often discussed, but do not have to
coincide with them.
Here we consider values of $T_{\rm r}$ in the range from $150\MeV$
to $3\times10^5\GeV$, which correspond to peak frequencies of GWs in the
ranges accessible to pulsar timing arrays \citep{Detweiler79, Hobbs+10,
NANOGrav2020} and space interferometers \citep{Caprini+16,
2017arXiv170200786A, 2021CmPhy...4...34T}.

As in \cite{Sharma+17} and SSS, we assume that $f$ is a function of
the scale factor $a$ with $f(a)\propto a^\alpha$ during inflation, and
$f(a)\propto a^{-\beta}$ during the post-inflationary matter-dominated
era, where $\alpha=2$ was fixed and $\beta$ is an exponent whose value
depends on $T_{\rm r}$.
The magnetic field becomes unstable and is rapidly amplified at large
length scales, provided the second derivative of $f$ with respect to
conformal time is positive.
This can be the case both for positive and negative exponents, i.e.,
both during and after inflation, but no longer in the radiation
dominated era, where $f=1$ must be obeyed for standard (conformally
invariant) electromagnetism to hold.

In contrast to BS, we now consider an additional term
$\Iota f^2 F_{\mu\nu} \tilde{F}^{\mu\nu}$ in the Lagrangian density,
where $\Iota$ is a constant and $\tilde{F}^{\mu\nu}$ is the dual
of the Faraday tensor.
The product is proportional to $\EE\cdot\BB$, where $\EE$
and $\BB$ are the electric and magnetic fields, respectively.
The term $\EE\cdot\BB$ is proportional to the rate of magnetic
helicity production.
The presence of such a term is common to many scenarios of helical
magnetogenesis, including the chiral magnetic effect \citep[CME;
see][]{Vilenkin80,joyce1997,BFR12,BFR15} and axion inflation
\citep{Barnaby+11, 1988PhRvD..37.2743T, fujita2015, adshead2016,
domcke2018, Domcke+20}.
In the case of magnetogenesis via axion inflation
\citep{carroll1992, adshead2016}, the helical term takes the form
$f_{\rm m}^{-1} \phi F_{\mu\nu} \tilde{F}^{\mu\nu}$, where $\phi$
represents the axion field and $f_{\rm m}$ is a mass scale associated
with the axion field.
In our model, $f(a)$ is constructed such that the model
avoids the aforementioned difficulties discussed in detail by
\cite{Sharma+17} and SSS.

As in BS, we employ the {\sc Pencil Code} \citep{JOSS} and apply it in
two separate steps.
In step~I, we solve the Maxwell and GW equations near the end of the
post-inflationary matter-dominated phase when the medium is still
electrically nonconducting and no fluid motions can be driven by the
Lorentz force.
Just like the (linearized) GW equation, the Maxwell equations are linear
and are advanced analytically between two subsequent times steps; see
Appendix~C of BS for details.
In step~II, when the conductivity has become large, we solve the standard
magnetohydrodynamic (MHD) equations.

The presence of the helical term proportional to
$\Iota$ leads to a difference in the growth rates between positively
and negatively polarized fields.
Fields with one of the two signs of helicities will therefore grow much
faster than the other.
Since there is enough time for the magnetic field to grow over many
orders of magnitude, it suffices to consider in step~I only fields
of one helicity.
This simplifies the computation somewhat.
In step~II, however, no such simplification is made.

In this paper, we work with conformal time $\eta$, which is related
to physical time $t$ through $\eta=\int\dd t/a(t)$.
By adopting appropriately scaled variables, we arrive at MHD equations
that are similar to those of standard MHD for a non-expanding Universe
\citep{BEO96}.
In step~I, during the post-inflationary matter-dominated era, the
effective equation of state is such that the scale factor increases
quadratically with conformal time (and like $t^{2/3}$ with physical time).
Conformal time is normalized such that it is unity at the beginning of
the subsequent radiation-dominated era.
Furthermore, the scale factor increases linearly with $\eta$ in the
radiation-dominated era.
We assume a spatially flat Universe and adopt the normalization of
\cite{RoperPol+20b,RoperPol+20}, where $a(\eta)=1$ at $\eta=1$ and
the mean radiative energy density is then also set to unity.

In \Sec{GovEqn}, we present the basic equations applied in steps~I and II.
Those for step~II are identical to the corresponding ones used in BS,
but the equations for step~I are different owing to the presence of the
magnetic helicity producing term proportional to $\Iota$.
We then present the results in \Sec{Results} and conclude in
\Sec{Conclusions}.
We adopt the Heaviside-Lorentz unit system and set the speed of light
equal to unity.

\section{The model}
\label{GovEqn}

\subsection{Polarization basis and governing equations}

Any vector field can be decomposed into an irrotational and
two vortical parts that are eigenfunctions of the curl operator
with positive and negative eigenvalues.
Here we employ the vector potential $\AAA$ in the Coulomb gauge,
$\nab\cdot\AAA=0$, so the irrotational part vanishes.
We then consider $\tildeAA(\eta,\kk)=\int {\AAA}(\eta,\xx)\,
e^{-\ii\kk\cdot\xx}\dd^3\xx$ in Fourier space, indicated by tildae,
as a function of conformal time $\eta$ and the wavevector $\kk$, and
write it as
\EQ
\tildeAA(\eta,\kk)
=\tildeA_+(\eta,\kk)\,\tildeee_+(\kk)
+\tildeA_-(\eta,\kk)\,\tildeee_-(\kk),
\EN
where
\EQ
\tildeee_\pm(\kk)=[\tildeee_1(\kk)\pm\ii\tildeee_2(\kk)]/\sqrt{2}\,\ii
\EN
is the polarization basis with $\ii\kk\times\tildeee_\pm=\pm
k\tildeee_\pm$, $k=|\kk|$ is the wavenumber and $\tildeee_1(\kk)$,
$\tildeee_2(\kk)$ represent units vectors orthogonal to $\kk$
and orthogonal to each other.
We assume an additional helical term in the EM Lagrangian density,
$f^2 F_{\mu\nu}(F^{\mu\nu}+\Iota\tilde{F}^{\mu\nu})$.
As in BS, we assume
\EQ
f(a)=a^{-\beta}\quad\mbox{with}\quad a=(\eta+1)^2/4
\label{fa_def}
\EN
being the scale factor during the post-inflationary matter-dominated era
with $-1<\eta\leq1$.
The evolution of the scaled vector potential,
$\tilde{\mathcal{A}}_\pm \equiv f \tildeA_\pm$, is then governed by
the equation \citep[SSS;][]{OF21}
\EQ
\tilde{\mathcal{A}}_\pm''+\left(k^2\pm2\Iota k\frac{f'}{f}
-\frac{f''}{f}\right)\tilde{\mathcal{A}}_\pm=0,
\label{dAk2dt2}
\EN
where primes denote $\eta$ derivatives, and
\EQ
\frac{f'}{f}=-\frac{2\beta}{\eta+1},\quad
\frac{f''}{f}=\frac{2\beta(2\beta+1)}{(\eta+1)^2}.
\label{d2f}
\EN
There are growing modes for $k<k_*(\eta)$, given by
\EQ
k_*(\eta)=2\beta\,\left(\Iota+\sqrt{1+\Iota^2+1/2\beta}\right)/(\eta+1),
\label{kpeak}
\EN
where we have considered the upper sign in \Eq{dAk2dt2}.
\EEq{kpeak} reduces to the expression given in Equation~(7)
of BS for $\Iota=0$.
For $\Iota=1$, we have $k_*(1)=\beta\,(1+\sqrt{2+1/2\beta})$.
For $\beta=7.3$, a particular case considered by BS, we have
$k_*(1)\approx18$ in the helical case when $\Iota=1$, which is more
than twice the value $k_*(1)\approx7.5$ for $\Iota=0$ used by BS
for the nonhelical case.
This shows that helicity broadens the range of unstable wavenumbers.
For $\Iota=-1$, we would have $k_*(1)\approx3.2$, but this is not
relevant in practice because the fastest growing mode would then
have opposite magnetic helicity,
and the results for $\Iota=1$ apply analogously.
Contrary to the case of nonhelical magnetogenesis ($\Iota=0$), where the
growth is fastest for $k=0$, it is now fastest for finite values of $k$.
In fact, as a function of $k$, the expression in round brackets in
\Eq{dAk2dt2} has an extremum for $k=2\beta\Iota/(\eta+1)$, and would
instead be at $k=0$ for $\Iota=0$.

As in BS, we also solve the linearized GW equations
\EQ
\tildeh_{+/\times}''
+\left(k^2-\frac{a''}{a}\right)\tildeh_{+/\times}
={6\over a}\,\tildeT_{+/\times}
\label{d2hdt2}
\EN
for the two polarization modes of the Fourier-transformed strain
$\tildeh_{+/\times}$.
As in \cite{RoperPol+20b,RoperPol+20}, we have made use of the
fact that the critical energy density at $\eta=1$ is unity.
The GWs are driven by the $+$ and $\times$ modes of the
traceless-transverse projected EM stress,
\EQ
{\TTT}_{ij}=f^2\,(B_i B_j+E_i E_j),
\label{TEB}
\EN
where $\EE=-\partial\AAA/\partial\eta$ and $\BB=\nab\times\AAA$ are the
electric and magnetic fields in real space.
We then compute $\tildeTTT_{ij}(\eta,\kk)=\int {\TTT}_{ij}(\eta,\xx)\,
e^{-\ii\kk\cdot\xx}\dd^3\xx$ in Fourier space, project out the
transverse-traceless part, and decompose the result into
$\tildeT_+$ and $\tildeT_\times$, which then enter in \Eq{d2hdt2};
see \cite{RoperPol+20b,RoperPol+20} for details.
In step~II, we solve the standard MHD equations with the usual
modifications for a radiation-dominated ultrarelativistic gas;
see also BS.
The bulk motions with velocity $\uu$
are nonrelativistic, but include second order terms in
the Lorentz factor \citep[see][for details]{BEO96,Bran+17}.
As stated before, the mean radiation energy density is set to unity
at $\eta=1$.
The new parameters in this step are the electric conductivity $\sigma$
and the kinematic viscosity $\nu$.
As in BS, we always assume the magnetic Prandtl number to be unity,
i.e., $\nu\sigma=1$.

\subsection{Diagnostics and initial conditions}

Important output diagnostics are energy spectra, $E_\lambda(\eta,k)$,
where $\lambda={\rm E}$, ${\rm M}$, ${\rm K}$, and ${\rm GW}$, for
electric, magnetic, kinetic, and GW energy spectra.
The symbols for the spectra are only used with these four subscripts
and are not to be confused with the components of the electric field
vector $\EE$.
The corresponding energy densities are defined as $k$ integrals over
these spectra, i.e., ${\cal E}_\lambda(\eta)=\int E_\lambda(\eta,k)\,\dd k$,
and are normalized such that $\EEEl=\bra{\EE^2}/2$, $\EEM=\bra{\BB^2}/2$, 
$\EEK=\bra{\uu^2}/2$, $\EEGW=\bra{h_+^2+h_\times^2}/6$.

We emphasize that $\EGW(k)$ denotes the GW energy density per linear
wavenumber interval, normalized to the radiation energy density at
$\eta=1$.
To obtain the GW energy density per logarithmic wavenumber interval,
normalized to the critical energy density today, one has to multiply
$k\EGW(k)$ by the dilution factor $(a_{\rm r}/a_0)^4(H_{\rm r}/H_0)^2$,
where the subscripts `r' and `0' refer to the scale factor $a$ and
the Hubble parameter $H$ at the end of reheating and today; see
\cite{RoperPol+20} for details regarding the normalization.
This leads to the quantity $h_0^2\OmGW(k)=1.6\times10^{-5}\,
(g_{\rm r}/100)\,k\EGW(k)$, where $g_{\rm r}$ is the number of
relativistic degrees of freedom at the beginning of the radiation
dominated era.

The simulations usually start at the initial time $\eta_{\rm ini}=-0.9$,
which implies $a(\eta_{\rm ini})=2.5\times10^{-3}$.
In some cases (Runs~C and D below), we used $\eta_{\rm ini}=-0.99$,
so that $a(\eta_{\rm ini})=2.5\times10^{-5}$.
As discussed in BS, the initial magnetic field has usually a
spectrum $\EM(k)\propto k^3$ for $k<\kpeak(\eta_{\rm ini})$.
The value of $\kpeak(\eta_{\rm ini})$ usually lies between the
smallest and largest wavenumbers in the computational domain, $k_1$
and $k_{\rm Ny}$, respectively, where $k_{\rm Ny}=k_1 n_{\rm mesh}/2$
is the Nyquist wavenumber and $n_{\rm mesh}$ is the number of
mesh points of the domain of size $2\pi/k_1$.
In this paper, we use $n_{\rm mesh}=512$ and we treat $k_1$ as an input
parameter that is usually chosen to be unity, but sometimes we also
consider smaller and larger values between 0.2 and 10, respectively.

The transition from step~I to step~II is discontinuous, as was already
discussed in BS.
This may be permissible when the change from zero conductivity to a
finite and large value occurs rapidly; see Appendix~D of BS.
In addition, while in step~II we have $f=1$, and therefore $f'=f''=0$,
the values of $f'/f$ and $f''/f$ at the end of step~I are small, but
finite, which can cause artifacts.
BS noted the occurrence of oscillations shortly after transitioning
to step~II, but the results presented for our GW spectra are always
averaged over the statistically steady state and are therefore
independent of the oscillations caused by the discontinuities of
these two ratios.
In the present case of helical magnetogenesis, there is also another
effect on the spectral slope of the GW energy density that will be
addressed below.

Let us emphasize at this point that in step~II, when $\sigma$ is large,
magnetic helicity, $\bra{\AAA\cdot\BB}$, is well conserved.
This is not the case in step~I, which is the reason why a helical
magnetic field can be produced.
Indeed, the magnetic helicity then grows at the same speed as the magnetic
energy grows.

\begin{figure*}\begin{center}
\includegraphics[width=.99\columnwidth]{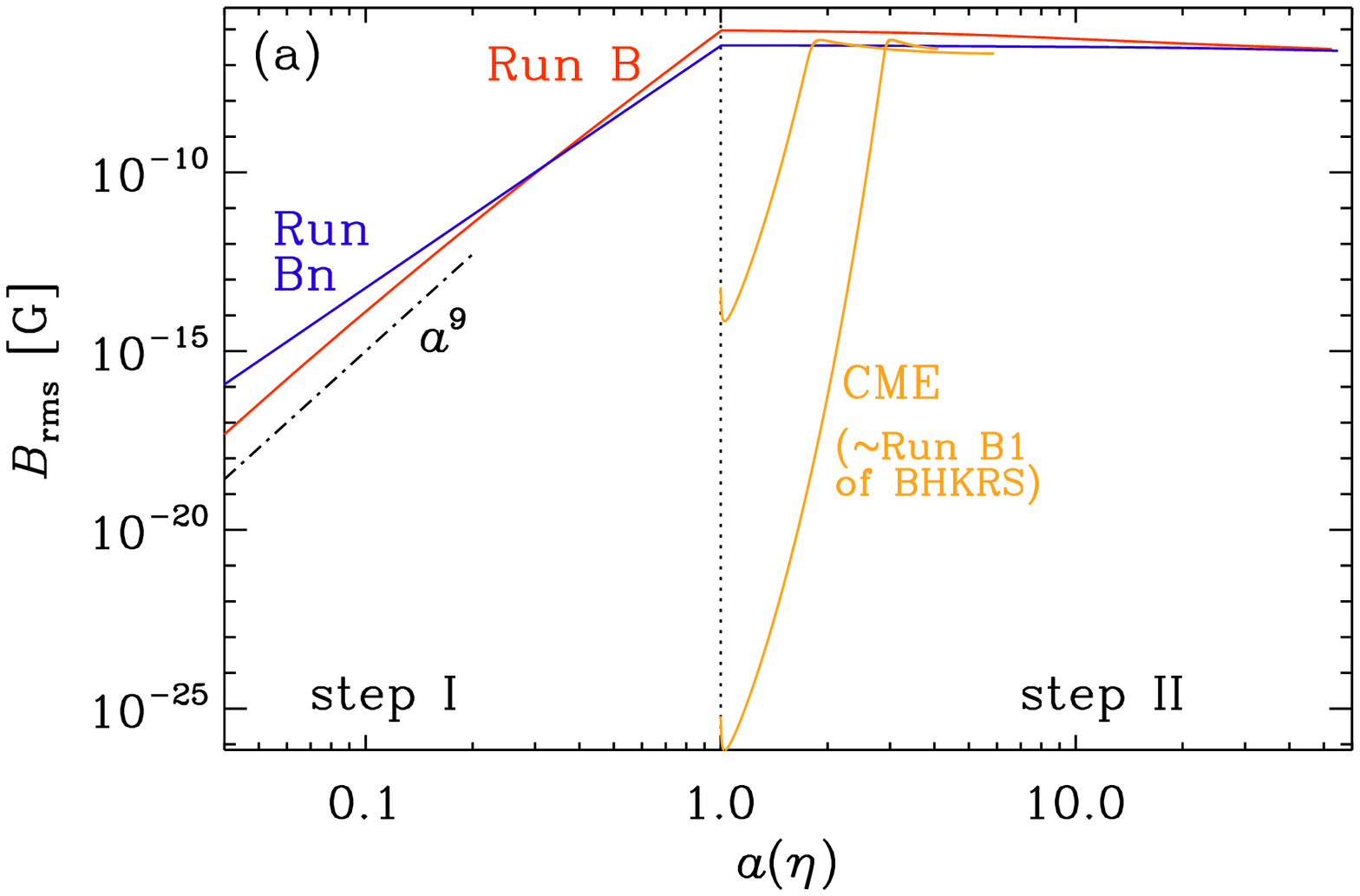}
\includegraphics[width=.99\columnwidth]{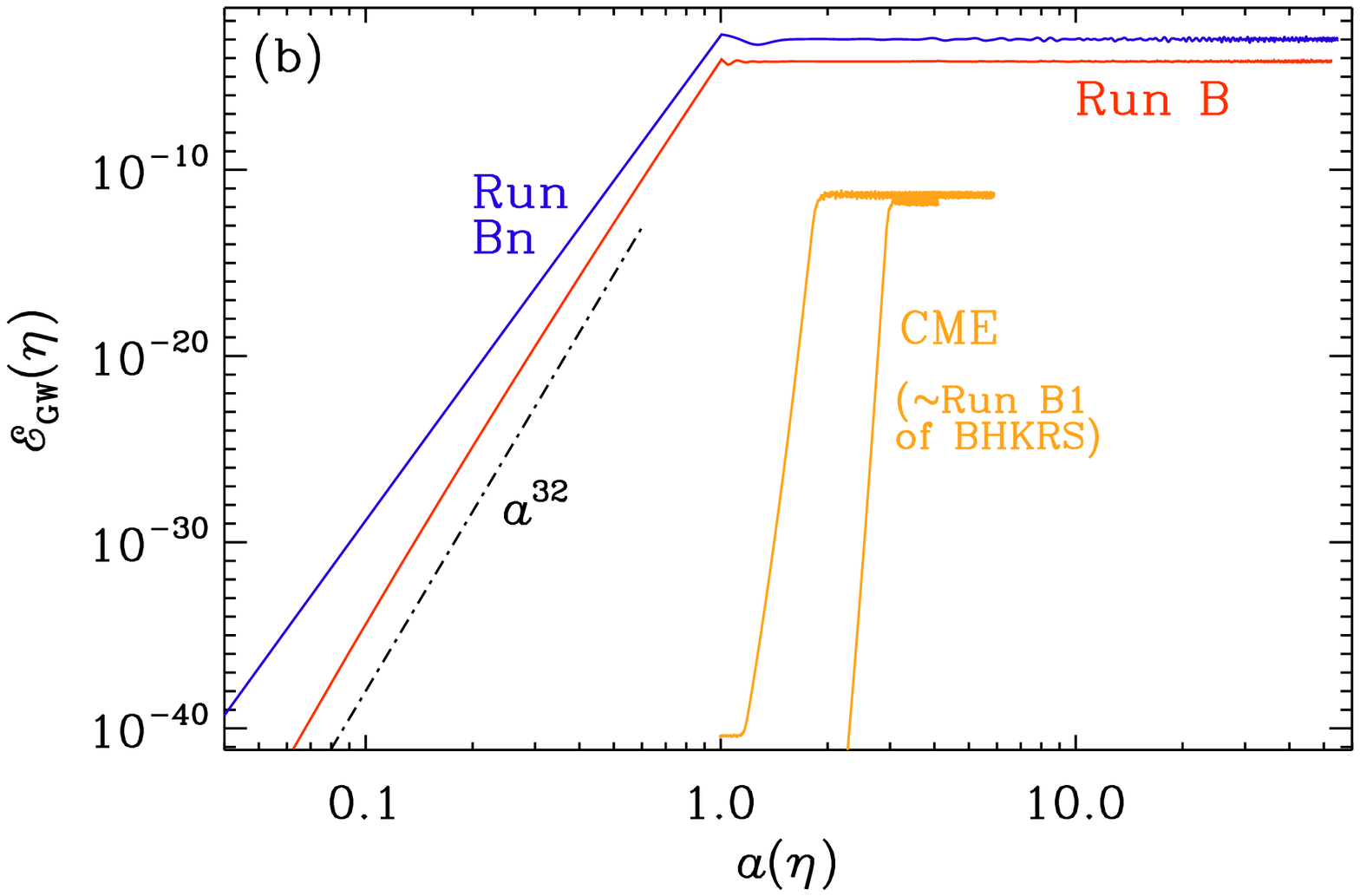}
\end{center}\caption{
Evolution of (a) $\Brms$ and (b) $\EEGW$ for Runs~B (red lines)
and Bn (blue lines), compared with two versions of Run~B1 of BHKRS
with different initial field strengths.
The two orange lines denote Run~B1 of BHKRS with the original and
a $10^{12}$ times weaker initial field.
Note that for the helical growth, the slopes change with $a(\eta)$,
which is a consequence of the helical term.
}\label{pcomp_befaftP}\end{figure*}

\subsection{Parameters of the magnetogenesis model}
\label{MagnetogenesisModel}

To avoid back-reaction and strong coupling problems of magnetogenesis
during inflation, SSS assumed the function $f$ to grow in a particular
fashion.
In the beginning, it grows as $a^{\alpha}$, starting from the value unity.
To recover the standard EM theory at the end of reheating, $f$ is
further assumed to continue evolving as $f \propto a^{-\beta}$ in the
post-inflationary era, which is assumed to be matter dominated.
The procedure to obtain the value of $\beta$ for a particular value of the
reheating temperature $\Tr$ is the same as explained in Appendix~A of BS.
The only difference lies in Equation~(A1) of BS, which is obtained by
demanding that the total EM energy density is a certain fraction $\EEEM$
of the background energy density at the end of the post-inflationary
matter-dominated era, will be different in the helical case.
Details are given in \App{RelationReheatingTemperature}.

In the model of SSS, $\alpha=2$ was chosen to have a
scale-invariant magnetic energy spectrum during inflation.
However, in the post-inflationary era, when $f$ decreases, the part
that provides a scale-invariant spectrum during inflation decays and the
next order term becomes dominant, giving an $\EM \propto k^3$ spectrum
in the superhorizon limit.
In this case, when $\alpha=2$, the maximum possible value
of the reheating temperature is approximately $50\GeV$.
This value is different from the value given by SSS, which was $4000\GeV$.
This difference is due to the fact that in SSS, the extra
amplification due to the presence of the helical term was not
considered in the post-inflationary matter-dominated era.

In BS, we focussed on two sets of runs---one for a reheating temperature
of around $100\GeV$ and another for $150\MeV$.
The corresponding values of $\beta$ where then 7.3 and 2.7, respectively.
We begin with similar choices of $\beta$ here, too.
It turns out that for $150\MeV$, the appropriate value is now $\beta=2.9$,
but for the standard scenario with $\alpha=2$, for the reasons explained
above, models for $100\GeV$ would not be allowed in the helical case,
because they would lead to strong backreaction, which forces us to choose
$\approx10\GeV$ instead.
In that case, the appropriate value would be $\beta=7.7$; see
\Tab{Tbeta2_reduced} for a summary of parameter combinations and
\App{RelationReheatingTemperature} for further details.
To facilitate comparison with BS, we have reduced the value of $\Tr$
to $8\GeV$, which then corresponds to $\beta=7.3$.

\begin{table}\caption{
$\beta$ for different values of $\Tr$.
}\begin{tabular}{crcccccccc} 
$\Tr$ [GeV] &$\alpha$&$\EEEM$& $\beta$&$g_r(\eta_*)$&$\EM(\eta_{\rm ini},k)$\\
\hline
$ 10$ &2&0.07 & 7.7&86&$\propto k^3$\\
$  8$ &2&0.01 & 7.3&86&$\propto k^3$\\
$0.15$&2&0.01 & 2.9&61.75&$\propto k^3$\\
$460$ &$-3$&0.01& 3 &106.75&$\propto k^{-1}$\\
$3 \times 10^5$&1&0.01& 1.7&106.75&$\propto k^5$\\
\label{Tbeta2_reduced}
\end{tabular}
\end{table}

In this paper, we also explore the possibility of
a smaller value of $\alpha$.
This allows for higher reheating temperature scales without having
any back-reaction problem in the post-inflation matter-dominated era.
For the case $\alpha=1$, the value of the reheating temperature is
$3 \times 10^{5}\GeV$ when the Hubble parameter during inflation is
$H_{\rm f}=10^{14}\GeV$ and the total EM energy density is $1\%$ of
the background energy density at the end of reheating.
These large values of $H_{\rm f}$ and $\Tr$ were not possible for
the case when $\alpha=2$.
This case is listed in the last row of \Tab{Tbeta2_reduced} along with
other relevant parameters.

We also consider the model of \cite{OF21}, where $f(a)\propto a^{-3}$
both during inflation and in the post-inflationary era, i.e.,
$\beta=3=-\alpha$.
In their model, the product $\beta\Iota$ was found to be $7.6$
so as to have maximum magnetic field strength for the case when
the total EM energy density is 1\% of the background energy density;
see Equation~(2.19) of \cite{OF21}.
This corresponds to $\Iota=2.5$.
In that case, the initial magnetic field had a scale-invariant spectrum
proportional to $k^{-1}$ in the superhorizon limit.

Quantum fluctuations alone would not introduce a preference of one sign
of helicity over the other, so therefore both ${\cal A}_+$
and ${\cal A}_-$ would grow at the same rate if $\Iota=0$.
However, if the magnetic field was fully helical to begin with,
only one of the two signs of helicity would grow,
i.e., either ${\cal A}_+$ or ${\cal A}_-$, so the field
might remain helical even though $\Iota=0$ and
both solutions would still be equally unstable.
In the following, we allow for such a possibility in some of our
simulations.

\begin{table*}\caption{
Summary of simulation parameters and properties.
}\begin{center}
\begin{tabular}{ccccccc|ccccc|cc}
Run & $\Tr$ [GeV] & $B_0$ & $\beta$ & $\Iota$ &
$\kpeak^{(1)}$ & $\nu$ & $\EEM$ & $\EEEM$ & $\EEM/\EEEM$ &
$\EEGW$ & $\hrms$ & $\qM$ & $\qEM$ \\
\hline
A&$0.15$&$5\times10^{-10}$&$2.9$&$1$&$7.2$&$1\times10^{-4}$&$0.012$&$0.023$&$0.51$&$1.2\times10^{-5}$&$9.1\times10^{-3}$&$2.1$&$1.07$\\
B&$  10$&$4\times10^{-24}$&$7.3$&$1$&$ 17$&$2\times10^{-4}$&$0.050$&$0.11$&$0.48$&$6.6\times10^{-5}$&$3.6\times10^{-3}$&$2.9$&$1.37$\\
Bn&$ 10$&$3\times10^{-18}$&$7.3$&$0$&$7.5$&$2\times10^{-4}$&$0.007$&$0.19$&$0.04$&$1.0\times10^{-3}$&$2.4\times10^{-2}$&$  32$&$1.30$\\
C&$460 $&$1\times10^{-27}$&$3.0$&$2.5$&$ 15$&$1\times10^{-4}$&$0.014$&$0.017$&$0.80$&$1.6\times10^{-6}$&$8.1\times10^{-4}$&$1.4$&$1.14$\\
D&$3\times10^{5}$&$5\times10^{-6}$&$1.7$&$1$&$4.3$&$5\times10^{-4}$&$0.016$&$0.025$&$0.64$&$8.5\times10^{-5}$&$7.6\times10^{-3}$&$2.5$&$1.58$\\
Dn&$3\times10^{5}$&$1\times10^{-3}$&$1.7$&$0$&$1.9$&$2\times10^{-4}$&$0.016$&$0.052$&$0.30$&$2.8\times10^{-3}$&$5.7\times10^{-2}$&$6.6$&$1.98$\\
\label{Tsummary}\end{tabular}
\end{center}
\end{table*}

\section{Results}
\label{Results}

\subsection{Growth of magnetic field and GW energy}

In \Fig{pcomp_befaftP}, we show the growth and subsequent decay of the
root-mean square (rms) magnetic field $\Brms$ during steps~I and II,
and compare with a simulation of nonhelical inflationary magnetic
field generation (similar to Run~B1 of BS).
The growth is still approximately algebraic, but, as expected, it is
now faster than in the nonhelical case.
This is caused by the extra amplification resulting from the helical
term proportional to $\Iota$.
This term is reminiscent of the CME, which causes, however, exponential
magnetic field amplification \citep{joyce1997}.
The CME has been invoked in the study of GW production from the resulting
magnetic field both analytically \citep{anand2018} and numerically
\citep[][hereafter BHKRS]{BHKRS21}.
The difference in the temporal growth of $\Brms$ and $\EEGW$ between
the CME and helical magnetogenesis is demonstrated in \Fig{pcomp_befaftP}.
Here we have also overplotted two versions of Run~B1 of BHKRS.

During the subsequent decay phase, $\Brms$ is approximately equally
large for both inflationary and CME runs.
This is just because of our choice of parameters.
However, owing to the smaller length scales on which the CME operates,
the corresponding GW energy is now much smaller than for inflationary
magnetogenesis.
On the other hand, we also see that the growth, being exponential, is
much faster for the CME runs than for both the helical and nonhelical
inflationary magnetogenesis models.
This implies that the CME can reach saturation with an arbitrarily weak
initial seed magnetic field.
The saturation amplitude does, however, depend on the assumed initial
imbalance of left- and right-handed fermions, and may, in reality,
be much smaller than what has been assumed in the models of BHKRS.
By contrast, the maximum field strength from inflationary magnetogenesis
is determined by demanding that the total EM energy density is some
fraction of the background energy density at the end of reheating so
that there is no back-reaction.

In \Tab{Tsummary}, we summarize quantitative aspects of our new runs,
Runs~A--D, as well as two nonhelical ones, Runs~Bn and Dn, where
$\Iota=0$.
We list the reheating temperature $\Tr$ in GeV, the amplitude parameter
$B_0$ for the initial magnetic field, the aforementioned parameters
$\beta$, $\Iota$, $\kpeak^{(1)}$, and $\nu$, as well as the output
parameters $\EEM$, $\EEEM\equiv\EEEl+\EEM$, the ratio $\EEM/\EEEM$, the
values of $\EEGW$ and the rms strain $\hrms=\bra{h_+^2+h_\times^2}^{1/2}$,
as well as two different efficiency parameters $\qM$ and $\qEM$,
defined below.

As in BS, varying the initial magnetic field strength $B_0$ always
resulted in a purely quadratic change of $\EEM$, and a quartic change
of $\EEGW$.
It therefore suffices to present, for each combination of parameters
$\beta$ and $\Iota$, only one value of $B_0$, typically such that
$\EEEM$ is roughly in the expected range of between 0.01 and 0.1.

\begin{figure*}\begin{center}
\includegraphics[width=1.8\columnwidth]{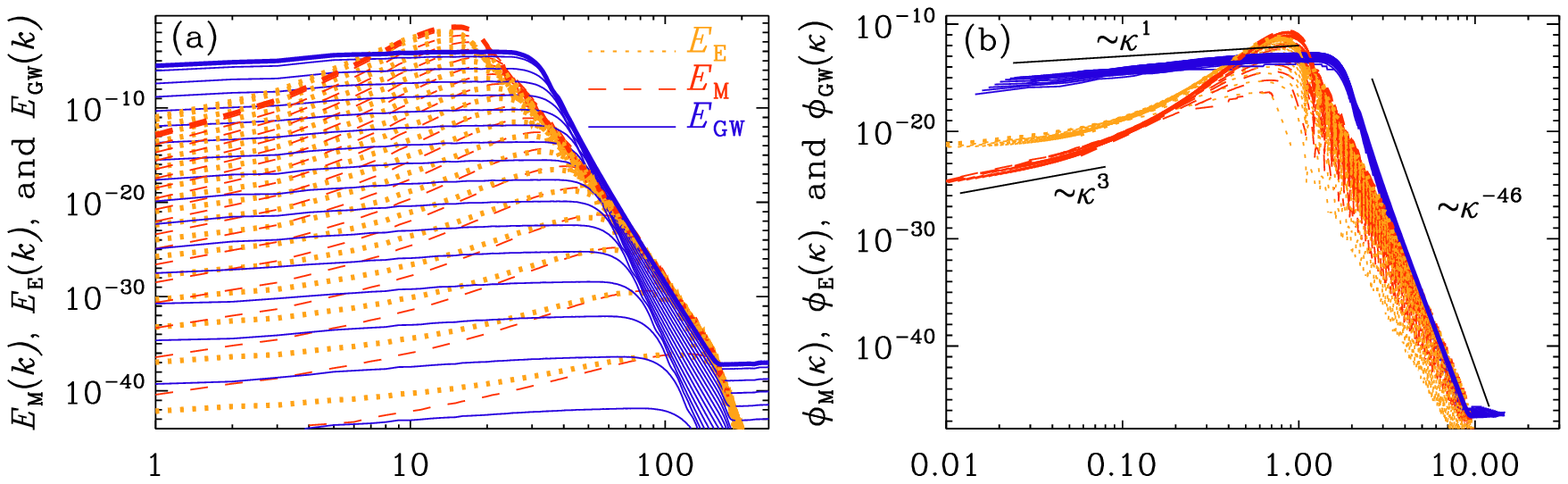}
\includegraphics[width=1.8\columnwidth]{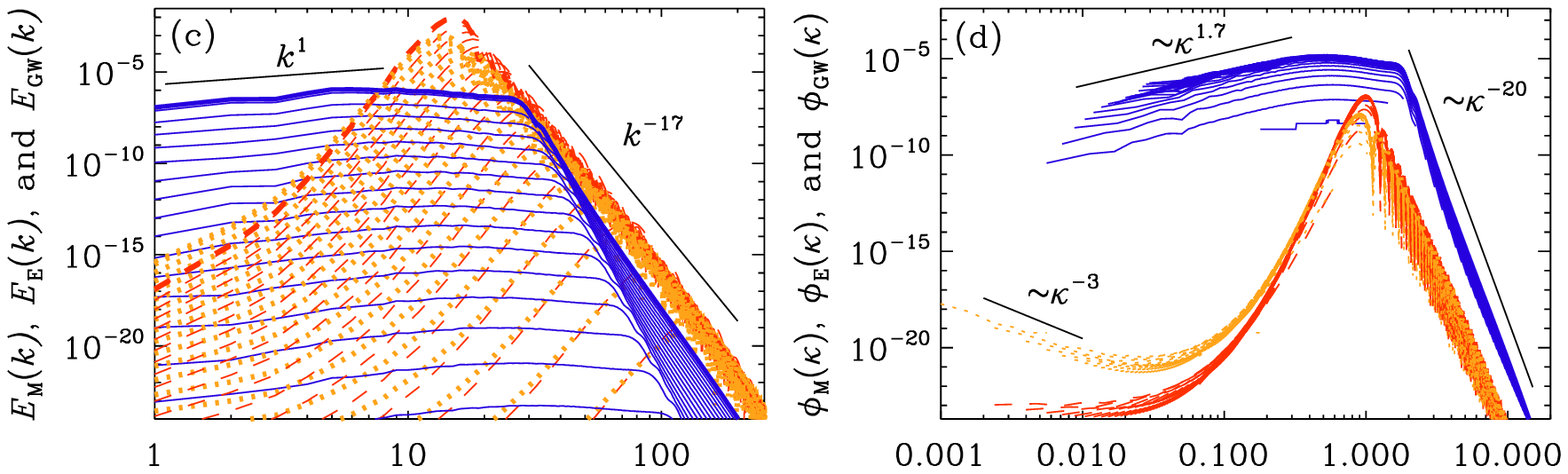}
\includegraphics[width=1.8\columnwidth]{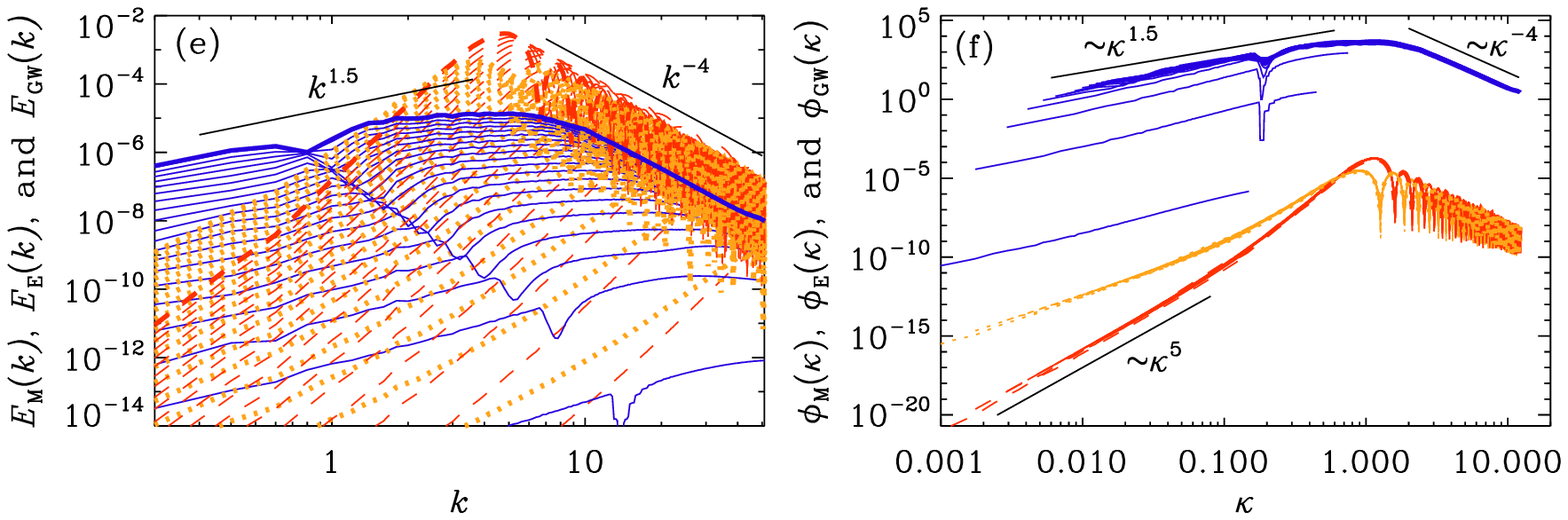}
\end{center}\caption{
$\EM(k)$ (red lines), $E_{\rm E}(k)$ (orange lines), and $\EGW(k)$
(blue lines) for (a) Run~B, (c) Run~C, and (e) Run~D, together
with the associated collapsed spectra 
$\phi_{\rm M}(\kappa)$ (red lines), $\phi_{\rm E}(\kappa)$ (orange lines),
and $\phi_{\rm GW}(\kappa)$ (blue lines) for
(b) Run~B, (d) Run~C, and (f) Run~D.
The spectral GW energy increases at a rate that is independent
of $k$, but the growth speed of $\EM(k)$ does depend on $k$.
}\label{rspec34P_P512b73b}\end{figure*}

\begin{figure*}\begin{center}
\includegraphics[width=\textwidth]{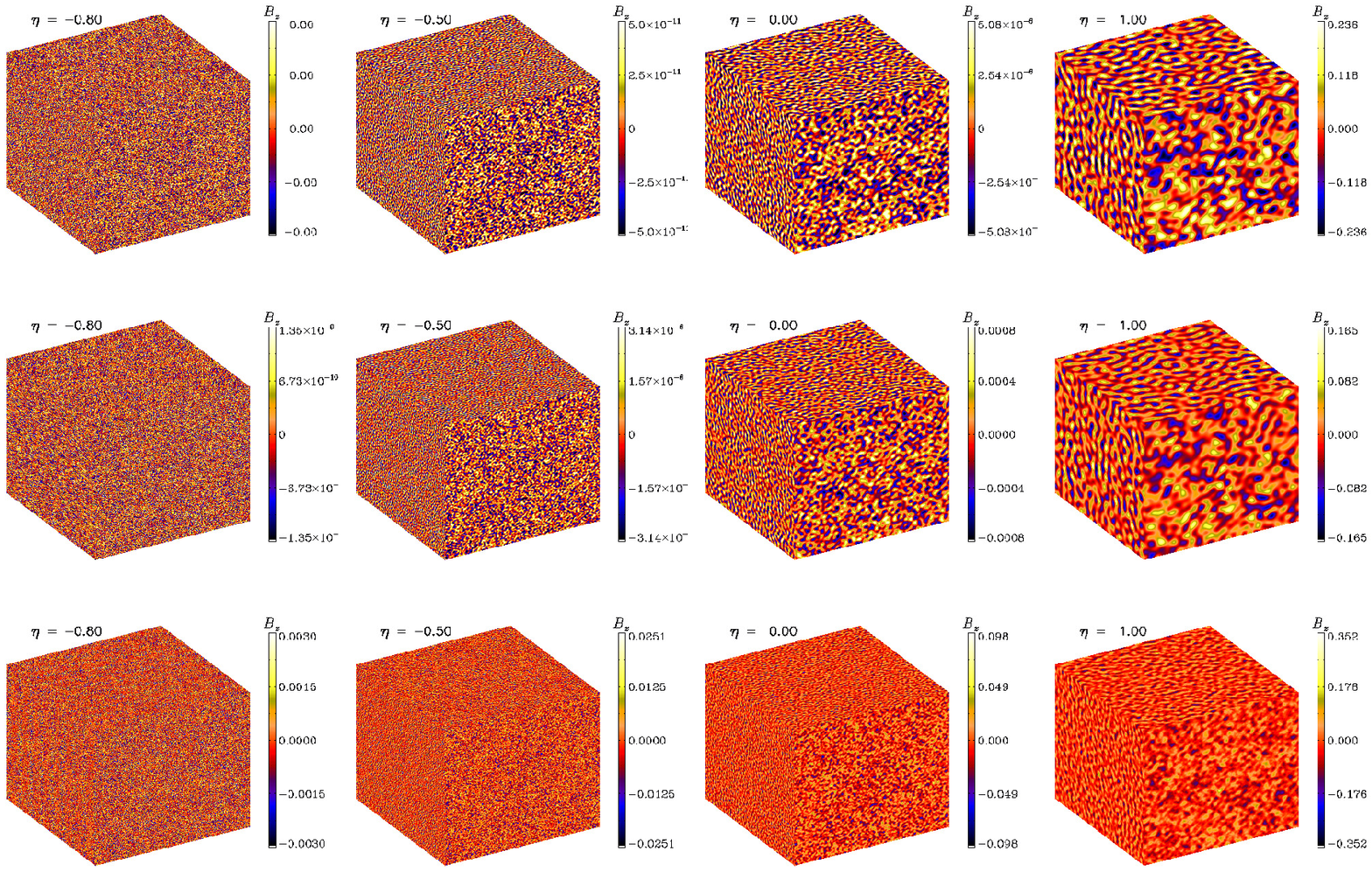}
\end{center}\caption{
Visualizations of $B_z$ for Runs~B (top), C (middle), and D (bottom)
on the periphery of the computational domain for $\eta=-0.8$, $-0.5$, $0$, and $1$
during step~I.
The color scale is symmetric about zero and adjusted with respect to
the instantaneous extrema.
}\label{ABC}\end{figure*}

Comparing helical with nonhelical runs for similar values of $\EEM$,
the GW energies and strains are smaller than in the earlier cases
without helicity (see also \Fig{pcomp_befaftP}).
This may suggest that GW production from helical inflationary
magnetogenesis is somewhat less efficient than for the nonhelical case.
However, while the values of $\EEM$ are the same,
the total EM energies, $\EEEM=\EEEl+\EEM$, are not.
In fact, we see that the ratio $\EEEl/\EEM$ is typically 0.3--0.5,
i.e., the electric energy contribution is subdominant during the
post-inflationary matter-dominated era.
For nonhelical magnetogenesis, by contrast, the electric energy is
dominant, typically with $\EEEl/\EEM=10$--$30$ for $\beta$ between
2.7 and 7.3.

\begin{figure}\begin{center}
\includegraphics[width=\columnwidth]{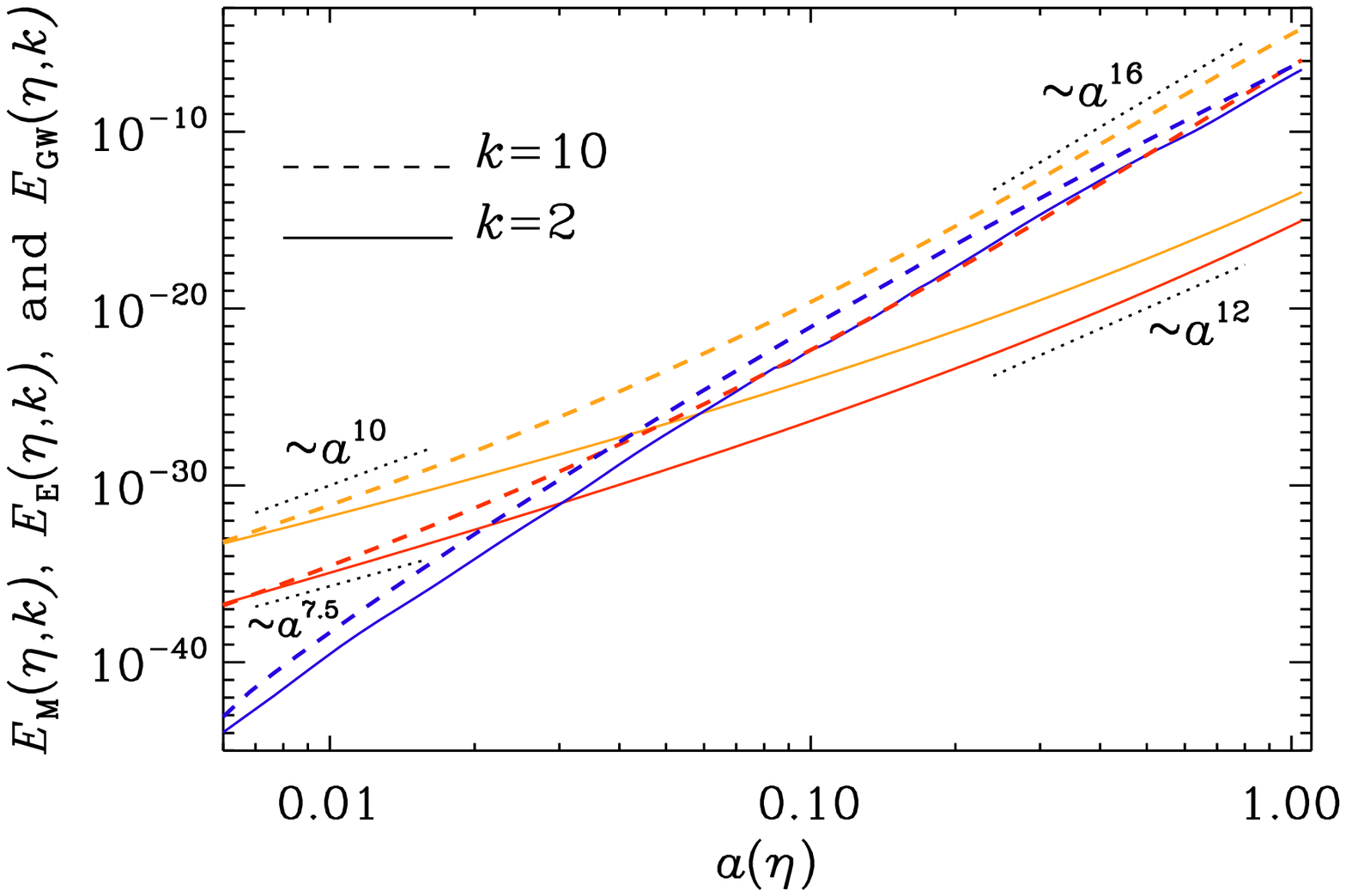}
\end{center}\caption{
Temporal dependence represented through $a(\eta)$ of spectral energies
at $k=2$ (solid lines) and $k=10$ (dashed lines) for Run~C
with $\EM(\eta,k)$ (red lines), $E_{\rm E}(\eta,k)$ (orange lines),
and $\EGW(\eta,k)$ (blue lines).
}\label{rspec34P_vs_eta_P512b73b}\end{figure}

As already noted, for fixed values of $\beta$ and $\Iota$, the different
values of $\EEM$, $\EEEM$, $\EEGW$, and $\hrms$
are directly related to the initial amplitude parameter $B_0$.
To compare runs with different parameters $\beta$ and $\Iota$,
we must therefore compute normalized efficiencies.
Earlier work \citep{RoperPol+20, BGKMRPS21} suggested that
$\EEGW=(\qM\EEM/k_{\rm c})^2$, where $\qM$ is the efficiency
and $k_{\rm c}$ is a characteristic wavenumber.
In analogy to their work, we now postulate an analogous
relation, but with $\EEEM$ instead of $\EEM$, i.e.,
\EQ
\EEGW=(\qEM\EEEM/k_{\rm c})^2,
\label{EEGW_from_qEM}
\EN
where $\qEM$ is a new efficiency parameter, and for $k_{\rm c}$
we always take the value $k_{\rm c}=\kpeak(1)$, just like in BS.

For nonhelical magnetogenesis, BS found that $\qM$ was proportional
to $\beta$.
Since $\kpeak(1)$ was also proportional $\beta$,
this meant that the effect of dividing by $\kpeak(1)$ was effectively
canceled, and that therefore a good scaling was obtained by just plotting
$\EEGW$ versus $\EEM^2$, suggesting that the $1/k_{\rm c}$ scaling may
not have been real.
However, our new results for helical magnetogenesis now show that this
is not the case for $\qEM$.
In fact, looking at \Tab{Tsummary}, where we present both $\qM$ and
$\qEM$, we see that $\qM$ shows significant variations ($1.4\la\qM\la32$),
while $\qEM$ changes comparatively little ($1.1\la\qEM\la1.6$).
This suggests that the GW energy is mainly governed by $\qEM$,
independently of or only weakly dependent on the value of $\beta$.

Among the four runs A--D, Runs~A and B are similar in that only the
value of $\beta$ is different.
For Runs~C and D, on the other hand, also the values of $\Iota$
and $\alpha$ were different.
In the following, therefore, we focus on presenting Runs~B--D
in more detail.

\subsection{Energy spectra}

Next, we compare Runs~B, C, and D by looking at the GW and magnetic
energy spectra for step~I during $-0.9\leq\eta\leq1$, where we also
compare with electric energy spectra.
As in BS, we try to collapse the spectra on top of each other by
plotting the functions
\EQ
\phi_\lambda(\kappa)=(\eta+1)^{-(p_\lambda+1)}E_\lambda(k,\eta),
\EN
where $\lambda={\rm E}$, ${\rm M}$, or ${\rm GW}$ for electric, magnetic,
and GW energies, respectively, 
$p_\lambda$ are exponents characterizing the speed of growth, for now
and
\EQ
\kappa(\eta)=k/k_*(\eta)
\EN
is a time-depended wavenumber where the EM energy spectra peak.
We show the result in \Fig{rspec34P_P512b73b}, where we
plot both $E_\lambda(k,\eta)$ and $\phi_\lambda(\kappa)$ for
Run~B in panels (a) and (b), Run~C in panels (c) and (d), and 
Run~D in panels (e) and (f).
We see that the tendency of the lines to collapse on top of each other
is better for the GW spectra than for the electric and magnetic spectra.
This shows that those latter two are not shape-invariant.
This is clearly different from the nonhelical case; see the corresponding
Figure~3 of BS.

Interestingly, except for the GW spectra, which show power law scalings
with $\EGW(k)\propto k$ for $k<2k_*(1)$ and $\EGW(k)\propto k^{-46}$
for $k>2k_*(1)$ (for Run~B), the EM spectra deviate from power law
scaling and show a more peaked spectrum for $k<k_*(1)$.
The growth is fastest in the model with $\beta=7.3$,
as is indicated by the spectra spanning about forty orders
of magnitude.
For Runs~C and D, the spectra are progressively more shallow.
For the GW spectrum of Run~D, there is a dip at $\kappa\approx0.17$
(and at decreasing values of $k$ as time increases).
This coincides with the wavenumber where $k^2=a''/a$ and thus, where
the solution to \Eq{d2hdt2} changes from oscillatory to temporally
growing behavior.
This feature is now so prominent, because the growth of the
magnetic field is now slower than before.

Visualizations of the magnetic field on the periphery of the computational
domain are shown in \Fig{ABC} for Runs~B--D.
We see that the typical length scales increase with time, but again
faster for Runs~B and C than for Run~D.

To study the temporal growth for specific values of $k$, we show in
\Fig{rspec34P_vs_eta_P512b73b} the dependencies of $\EEl(\eta,k)$,
$\EM(\eta,k)$, and $\EGW(\eta,k)$ separately for $k=2$ and $10$ for
Run~C, where the departure from shape-invariant behavior appears to be
the strongest.
We clearly see that the growth of $\EGW(\eta,k)$ is the same for all
values of $k$.
This is in agreement with the visual impression from
\Fig{rspec34P_P512b73b}.
It is also the same at early and late times.
This is not the case for the electric and magnetic spectra, where we have
a growth proportional to $a^{7.5}$ for $k=2$ and small values of $a$,
but a faster growth $\propto a^{20}$ for $k=10$ and $a(\eta)>0.1$.

When the mode corresponding to a certain wavenumber $k$ is well outside
the horizon, the $f''/f$ term within the round brackets of \Eq{dAk2dt2}
dominates over the other two terms, and the amplitude of the mode grows
in time.
Once the mode is about to enter the horizon, the second term also comes
into the picture and further enhances the growth rate for $\Iota=1$.
This behavior is shown in \Fig{rspec34P_vs_eta_P512b73b}.

\begin{figure}\begin{center}
\includegraphics[width=\columnwidth]{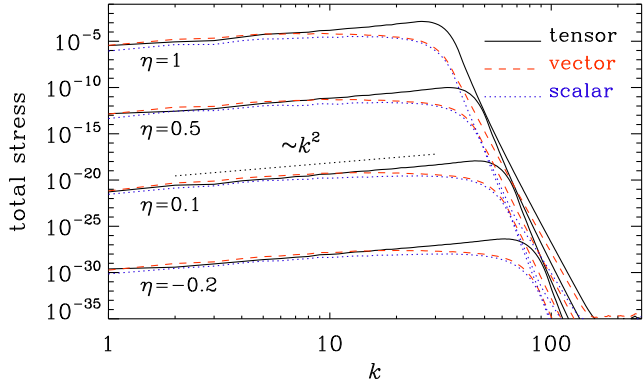}
\end{center}\caption{
Spectra of the total stress at $\eta=-0.2$, $0.1$, $0.5$, and $1$,
decomposed into tensor (solid black), vector (dashed red), and
scalar modes (dotted blue) for Run~B of \Fig{rspec34P_P512b73b}.
}\label{pSCL2_P512b73b}\end{figure}

To understand the nearly shape-invariant scaling of $\EGW(\eta,k)$,
it is important to look at spectra of the stress.
This is done in \Fig{pSCL2_P512b73b},
where we show spectra of the stress, decomposed into tensor, vector,
and scalar modes \citep{Mukhanov+92}.
The tensor mode is the transverse-traceless contribution to the
stress, while the vector and scalar modes are composed of vortical and
irrotational constituents, respectively; see \cite{BGKMRPS21} for such
a decomposition of data from earlier GW simulations.
We see that at all times during step~I, the scalar and vector modes
are subdominant.
In particular the peak of the stress spectrum is to a large fraction
composed of the tensor mode only.
As expected from the work of \cite{BB20}, its spectrum follows
a $k^2$ subrange to high precision.

\begin{figure*}\begin{center}
\includegraphics[width=1.8\columnwidth]{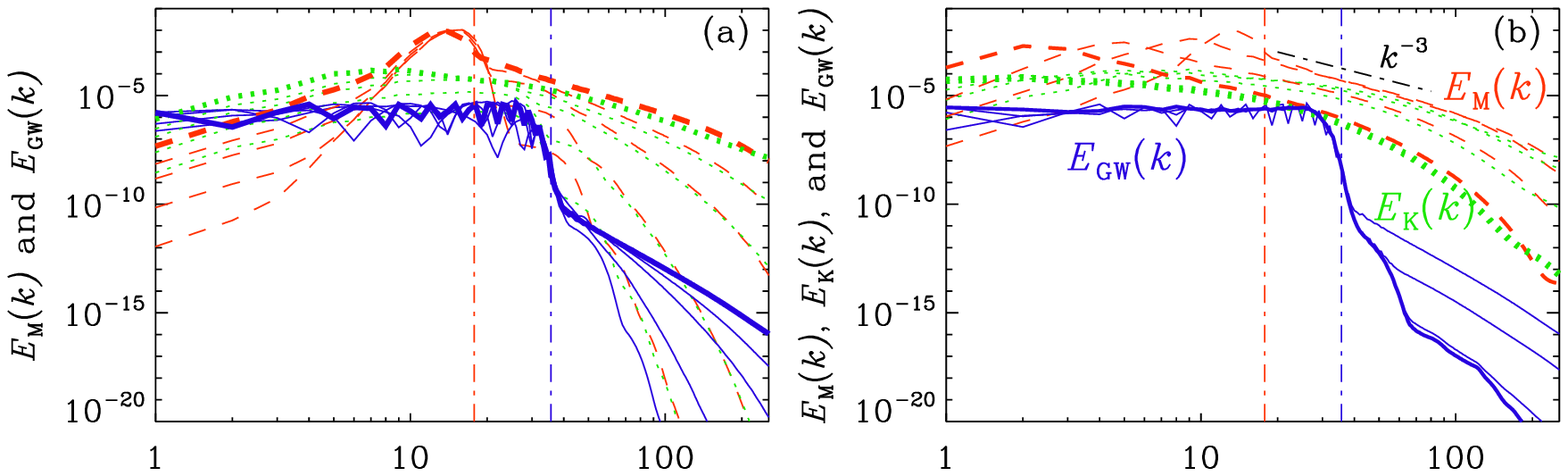}
\includegraphics[width=1.8\columnwidth]{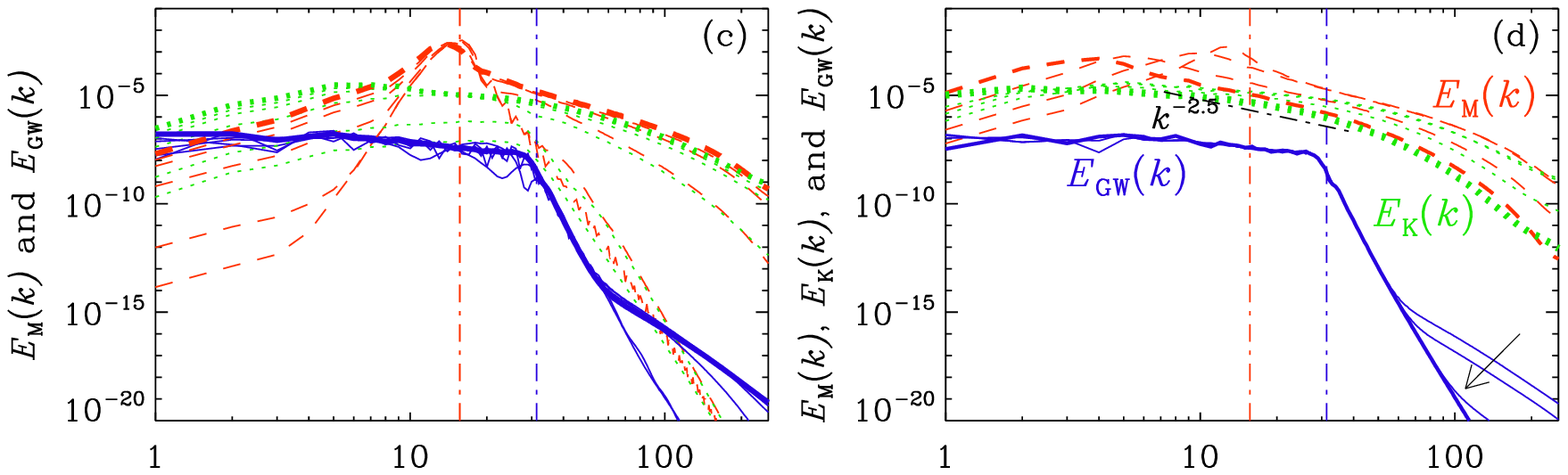}
\includegraphics[width=1.8\columnwidth]{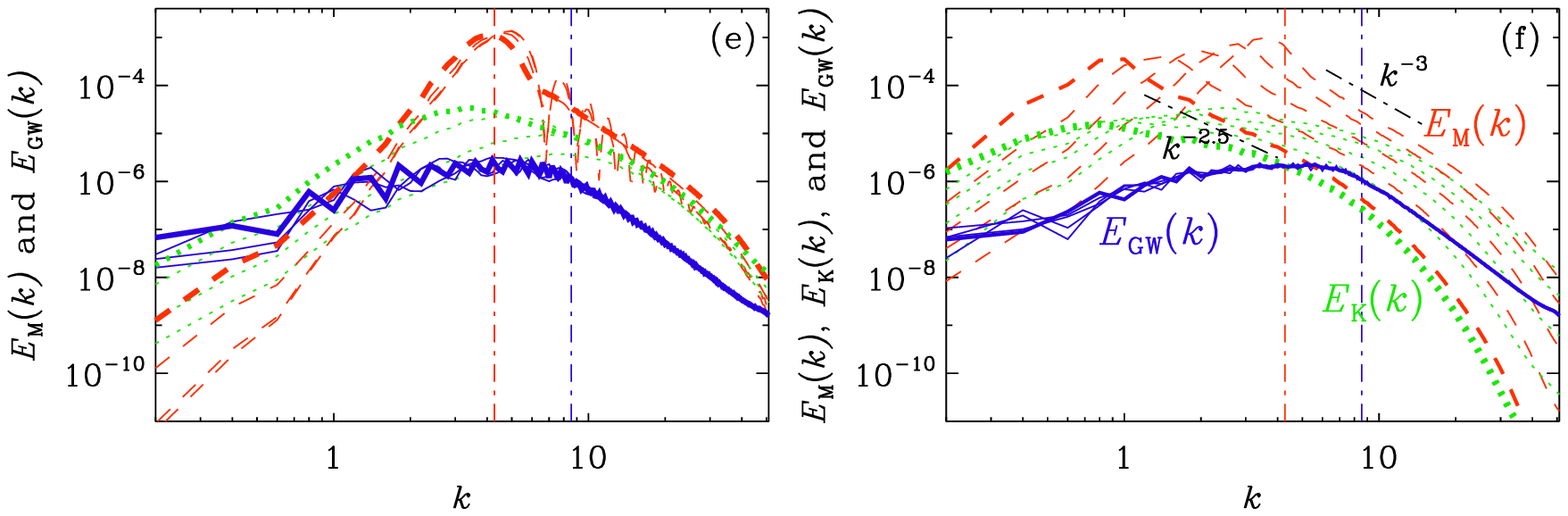}
\end{center}\caption{
Early times in the beginning of the radiation-dominated phase
for (a) Run~B ($\eta=1.06$, 1.2, 1.4, 1.6, and 2.1),
(c) Run~C ($\eta=1.06$, 1.9, 2.7, 3.3, and 4.1), and
(e) Run~D ($\eta=1.6$, 2.1, 3.6, and 6.1).
$\EM(k)$, $\EK(k)$, and $\EGW(k)$ are shown as dashed red,
dotted green, and solid blue lines, respectively.
The last times are shown as thick lines.
Later times are shown separately for
(b) Run~B ($\eta=2$, 6, 16, and 52),
(d) Run~C ($\eta=11$, 26, and 52), and
(f) Run~D ($\eta=11$, 26, 51, 101, and 213).
The red and blue vertical dashed-dotted lines goes through
$k_*(1)$ and $2k_*(1)$, respectively.
Again, thick lines denote the last time.
The arrow in panel (d) highlights the sense of time,
where $\EGW(k)$ declines at large values of $k$.
}\label{rspec3P_select_P512b73b_MHD}\end{figure*}

Comparing the different models, we see that for $\kappa\ll1$,
we reproduce the initial scalings $\phi_{\rm M}\propto\kappa^3$
for Run~B and $\propto\kappa^5$ for Run~D, with a shallower scaling
by a factor $\kappa^2$ for the electric fields, in particular the
$\phi_{\rm E}\propto\kappa^{-3}$ scaling for Run~C.
For $\kappa\gg1$, we have a progressively shallower decline
$\propto\kappa^{-46}$, $\kappa^{-20}$, and $\kappa^{-4}$
as we go from Run~B to Runs~C and D.

\subsection{Spectra in step~II}

In step~II, a velocity field emerges, driven by the Lorentz force.
This causes the magnetic field to develop small-scale structure,
as can be seen from \Figp{rspec3P_select_P512b73b_MHD}{a}.
This leads to a turbulent cascade that has here a spectrum
proportional to $k^{-3}$ for large $k$;
see \Figp{rspec3P_select_P512b73b_MHD}{b}.
Contrary to BS, the new GW spectrum now shows a flat power law
scaling for $k<2k_*(1)$ with $\EGW(k)\propto k^0$, i.e.
$k\EGW(k)\propto k^1$.
Such a scaling was already found by \cite{RoperPol+20}.
The reason for this lies in the direct correspondence with the relevant
magnetic stress for the blue-tilted magnetic energy spectrum, where
$\EM(k)$ has an increasing slope with an exponent larger than two,
which corresponds to a white noise spectrum.
In that case, this stress itself always has a white noise spectrum and
cannot be steeper than that.
This was shown by \cite{BB20}, who just considered the stress spectrum
and ignored temporal aspects, i.e., they did not consider solutions to
the GW equation.

\begin{figure*}\begin{center}
\includegraphics[width=2\columnwidth]{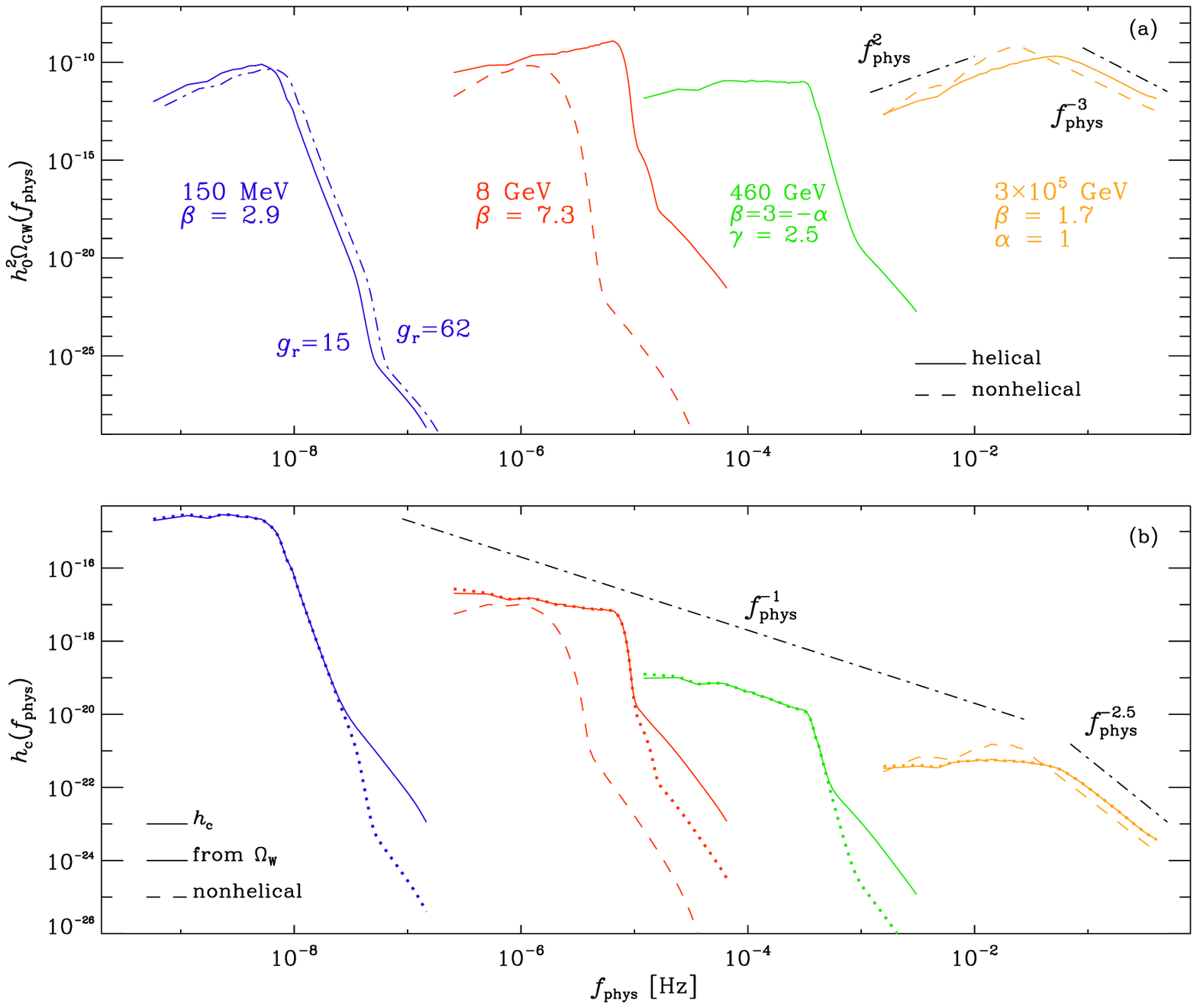}
\end{center}\caption{
(a) $h_0^2\Omega_{\rm GW}(f_{\rm phys})$ and (b) $h_c(f_{\rm phys})$ for
Runs~A--D $T_{\rm r}$ ranging from $150\MeV$ to $3\times10^5\GeV$.
In (a), dashed lines denote nonhelical runs and dashed-dotted show
the result for $g_{\rm r}=62$.
In (b), the dotted lines denote
$1.26\times10^{-18}\sqrt{h_0^2\OmGW}\,(1\Hz/f_{\rm phys})$ \citep{Mag00}.
}\label{pspecmP2}\end{figure*}

As in BS, the GW spectrum shows a marked drop by about six orders of
magnitude for Run~B, which is slightly more than what was found in BS.
We return to this in \Sec{ObservableSpectra}, but
we note at this point that for $k\gg2\kpeak(1)$ in Runs~B and C,
the spectral GW energy beyond the drop,
which is very small already, becomes even smaller as time goes on.
This is indicated by the arrow in \Figp{rspec3P_select_P512b73b_MHD}{d}.
Eventually, the spectrum settles at a level close to the fat blue lines
in \Fig{rspec3P_select_P512b73b_MHD}, which marks the last time.
Furthermore, at late times, \Figp{rspec3P_select_P512b73b_MHD}{b}
shows clear inverse cascading with the peak of the magnetic
spectra traveling towards smaller $k$; see the red dashed lines in
\Fig{rspec3P_select_P512b73b_MHD}.
The height of the peak is expected to stay unchanged \citep{BK17}, but
our present runs show a small decline with time.
This is predominantly a consequence of the conductivity still not being
high enough.
Larger conductivity would require larger numerical resolution, which
would begin to pose computational memory problems.

In step~II, the GW spectrum is now fairly flat,
$\EGW\propto k^0$ for Runs~B and C, and with a slight rise
$\propto k$ for Run~D.
Therefore, the GW energy per logarithmic wavenumber interval, normalized
by the critical energy density for a spatially flat universe, is
$\OmGW\propto k\EGW\propto k^1$ for Run~B, and perhaps even slightly
shallower for Run~ C, and $\propto k^2$ for Run~D.
Thus, as already seen in many earlier numerical simulations of
turbulence-driven GWs \citep[][BHKRS]{RoperPol+20}, this is shallower than
the previously expected $k^3$ scaling \citep{Gogo+07,OF21}.
In the present case, during the onset of MHD turbulence, the spectrum
has changed from a $k^1$ spectrum to a $k^0$ spectrum.
As explained in Appendix~F of BS, this is associated with the
discontinuous behavior of $f'/f$ and $f''/f$.
They concluded that the change from a $k^1$ spectrum to $k^0$ 
occurs when the growth of EM energy has stopped.
This is at the same time when $f'=f''=0$, but it is not a direct
consequence of the discontinuity at $\eta=1$ and therefore not an
artifact.

We see clear inverse cascading in the magnetic energy spectra
with the peak of the spectrum moving toward smaller $k$.
This has been investigated in detail in many earlier papers
\citep{Hat84,BM99}; see \cite{BK17} for a demonstration of the
self-similarity of the magnetic energy spectra.
The conservation of mean magnetic helicity density,
$\bra{\AAA\cdot\BB}$, implies a growth of the correlation length
and a corresponding decay of the mean magnetic energy density such
that $\bra{\AAA\cdot\BB}\approx\pm\Brms^2\xiM\approx\const$ for fully
helical turbulence, where the two signs apply to positive and negative
magnetically helicities, respectively.

\subsection{Observable spectra}
\label{ObservableSpectra}

In \Fig{pspecmP2}, we show the final spectra of $\OmGW$ and $\hc$ versus
temporal frequency $f_{\rm phys}=k H_*/2\pi a_0$ for the present time.
The frequency $f_{\rm phys}$ is not to be confused with the function
$f(a)$, defined in \Eq{fa_def}, which does not carry any subscript.
Both the strain and the energy spectra are scaled for the corresponding
values of $T_{\rm r}$ between $150\MeV$ and $3\times10^5\GeV$.
We have indicated spectra for the nonhelical case as dashed lines.

The spectra in \Fig{pspecmP2} show different shapes of the $\OmGW$
spectra for helical and nonhelical runs.
This may, to some extent, be caused by the larger values of $\kpeak(1)$
in these helical runs.
Furthermore, the drop beyond the peak is stronger in the helical case.
This was also found in previous simulations \citep{RoperPol+20,
2021arXiv210212428B}, and may be related to the presence of a weaker
forward cascade in favor of a stronger inverse cascade in helical
turbulence \citep{PFL76}.
Note also that for Run~B with the largest value of $\beta$,
the change from the scaling $\OmGW\propto f_{\rm phys}$
is much sharper in the case with helicity than without, where the
spectra are much rounder.

In the model with $\Tr=150\MeV$, we compare the GW spectra generated both
before and after the QCD phase transition, where $g_{\rm r}$ changes by
a factor of about four from 62 to about 15.
This leads to a drop in frequency by a factor $\propto g_{\rm r}^{1/2}$ of
about two and in an increase in GW energy by a factor $\propto g_{\rm r}^{1/3}$
of about $1.6$.

\begin{figure*}\begin{center}
\includegraphics[width=\textwidth]{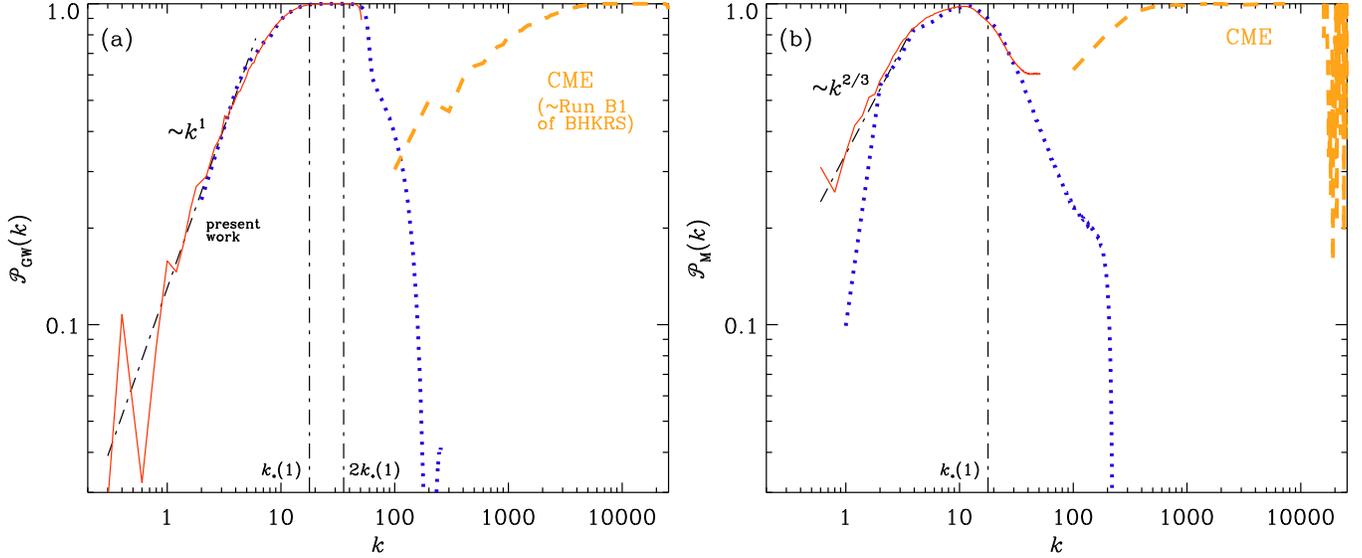}
\end{center}\caption{
(a) ${\cal P}_{\rm GW}(k)$ and (b) ${\cal P}_{\rm M}(k)$
for Run~B (with $k_1=1$; blue solid line)
and a corresponding run with $k_1=0.2$ (red dashed-dotted line),
as well as for Run~B1 of BHKRS (orange dashed line).
The vertical dashed-dotted lines mark the positions of
$\kpeak(1)$ in (a) and (b) and of $2\kpeak(1)$ in (a).
}\label{ppol_comp}\end{figure*}

We see that the high $\Tr$ model is different from the
other models with lower $\Tr$ in several respects.
The drop in GW energy above the maximum is now absent and the
inertial range slope is no longer $\propto f_{\rm phys}$,
but to $\propto f_{\rm phys}^2$.
This is mainly caused by the small value of $\beta$, which
results in a slower growth.
At the same time, the spectral peak at $\kpeak(\eta)$ still moves
to smaller values as before.
This causes the slope for $k>2\kpeak(1)$ to be shallower than in
the other models with larger values of $\beta$.
The slope is then also inherited in step~II, and it is then not
much affected any more by the emerging turbulence.

The model of \cite{OF21} with $\Tr=460\GeV$ corresponds to our Run~D.
They also studied GW production, but they did not include the turbulent
phase after reheating.
Comparing our \Fig{pspecmP2} with Figure~5 of \cite{OF21}, we see that
the peak values are slightly different.
Our spectral peak is at approximately $h_0^2\OmGW\approx10^{-11}$,
while their peak value without the $h_0^2$ factor is $\OmGW\approx10^{-12}$.
Furthermore, as we saw already from \Fig{rspec3P_select_P512b73b_MHD}, the
slope of $\EGW(k)$ was slightly negative close to the peak.
Therefore, the $\OmGW(k)\propto k\EGW(k)$ is now nearly flat.
This is quite different from Figure~5 of \cite{OF21}, which had
a clear $\OmGW(k)\propto k^3$ range below the peak.
The frequency corresponding to the peak is also slightly different, but this
is to some extent explained by their frequency lacking a $2\pi$ factor.

\subsection{Circular polarization}

In \Figp{ppol_comp}{a}, we plot the time-averaged fractional circular polarization
spectrum of GWs, ${\cal P}_{\rm GW}(k)$, for Run~B.
It is defined as \citep[see Equation~B.17 of][]{RoperPol+20b}
\EQ
\!\!{\cal P}_{\rm GW}(k) = \!\left.\int\!2\,\Imag\,\tildeh_+\tildeh_\times^*\,k^2\dd\Omega_k
\right/\!\!\int\!\left(|\tildeh_+|^2+\tildeh_\times|^2\right) k^2\dd\Omega_k.
\EN
In \Figp{ppol_comp}{b}, we show the fractional magnetic helicity spectrum,
\EQ
{\cal P}_{\rm M}(k) = k\HM(k)/2\EM(k),
\EN
where $\HM(k)$ is the magnetic helicity spectrum, normalized such that
$\int\HM(k)\,\dd k=\bra{\AAA\cdot\BB}$.
Unlike the GW spectrum, which is statistically stationary and we can
take a long-term average, the magnetic field develops a forward cascade
and decays at the same time.
During that time, the kinetic energy density has a maximum, which marks
the moment when the turbulent cascade has developed.
We have therefore decided to take a short-term average of the magnetic
helicity and energy spectra around the time when the kinetic energy
density is within about 70\% of its maximum value.

We also compare with the corresponding spectrum from Run~B1 of BHKRS
with CME (not to be confused with Run~B1 of BS).
Except for a hundredfold shift toward larger $k$, the shapes of
${\cal P}_{\rm GW}(k)$ are similar in that both have a plateau with
${\cal P}_{\rm GW}(k)\approx1$ and a similar decline toward smaller
values of $k$.

Toward larger values of $k$, we see a drop in ${\cal P}_{\rm GW}(k)$
that is superficially similar to the drop in GW energy---at least for
the present runs.
In the runs driven by the CME, such a drop is absent.
However, the drop in the GW energy spectra for large $k$ is probably
not related to the drop seen in the polarization spectra, where it
appears for a larger $k$ value of nearly $4\kpeak(1)$.
Furthermore, at about $k=\kpeak(1)$, we rather see that ${\cal P}_{\rm GW}(k)$
declines toward {\em smaller} $k$ values, i.e., for $k<2\kpeak(1)$.

\begin{table*}\caption{
Present day values for Runs~A--D using
parameters from \Tab{Tsummary} as input,
assuming always $\EEEM=0.01$.
}\begin{tabular}{ccccccccc} 
Run & $\Tr$ [GeV] & $\eta_{\rm eq}$ & $\xiM^*$ [Mpc] & $\xiM^{\rm eq}$ [Mpc] &
$\Brms^*$ [G] & $\Brms^{\rm eq}$ [G] & $\EEGW$ & $h_0^2\OmGW$ \\
\hline
A&$0.15$&$3.8\times10^{8}$&$5.8\times10^{-8}$&$3.0\times10^{-2}$&$3.0\times10^{-7}$&$4.2\times10^{-10}$&$2.2\times10^{-6}$&$4.3\times10^{-11}$\\
B&$  10$&$2.8\times10^{10}$&$3.2\times10^{-10}$&$2.9\times10^{-3}$&$2.9\times10^{-7}$&$9.6\times10^{-11}$&$5.3\times10^{-7}$&$9.2\times10^{-12}$\\
C&$ 460$&$1.4\times10^{12}$&$8.0\times10^{-12}$&$9.9\times10^{-4}$&$3.8\times10^{-7}$&$3.4\times10^{-11}$&$5.3\times10^{-7}$&$8.5\times10^{-12}$\\
D&$3\times10^{5}$&$9.0\times10^{14}$&$4.5\times10^{-14}$&$4.2\times10^{-4}$&$3.4\times10^{-7}$&$3.5\times10^{-12}$&$1.4\times10^{-5}$&$2.2\times10^{-10}$\\
\label{Ttoday}
\end{tabular}
\end{table*}

We have also confirmed that the decline below $k=\kpeak(1)$
is not related to the finite domain size.
We have also performed a simulation with a five times larger domain,
where $k_1=0.2$ instead of $k_1=1$.
By comparing these two runs, we recovered essentially the
same ${\cal P}_{\rm GW}(k)$ profile.
This is shown in \Fig{ppol_comp} as the red dashed line,
which agrees with the blue one for $k_1=1$ for not too small
$k$ values.
In particular, we see that there is evidence for a linear
scaling of the fractional polarization, i.e., 
${\cal P}_{\rm GW}(k)\propto k$.

Comparing with the fractional magnetic helicity spectrum,
${\cal P}_{\rm M}(k)$, we see that it also declines toward
smaller $k$, but this happens more slowly.
In fact, for Run~B, where ${\cal P}_{\rm GW}(k)$ already
declines, ${\cal P}_{\rm M}(k)$ is just reaching its maximum.
For larger values of $k$, we see that ${\cal P}_{\rm M}(k)$
already declines for Run~B while ${\cal P}_{\rm GW}(k)$
is still at its plateau.
However, for the CME runs, no decline in ${\cal P}_{\rm M}(k)$
is seen.

\subsection{Present day values}

The values of $\EEM$ listed in \Tab{Tsummary} gave the magnetic energy
fraction of the radiation energy at $\eta=1$.
To obtain the comoving rms magnetic field in gauss, we set
$\Brms^2/8\pi=\EEM\,(\pi^2 g_0/30)\,(\kB T_0)^4/(\hbar c)^3$, where
$g_0=3.94$ and $T_0=2.7\K$ is the present day temperature, $\kB$ is
the Boltzmann constant, and $\hbar$ is the reduced Planck constant.
By using $\EEEM=0.01$ in all cases, we can compute $\EEM$ by taking
the $\EEM/\EEEM$ ratios from \Tab{Tsummary} for Runs~A--D.
Likewise, we use \Eq{EEGW_from_qEM} with the $\qEM$ values listed in
that table and compute $h_0^2\OmGW$ from $\EEGW$ by multiplying with
the appropriate dilution factor.

At $\eta=1$, the typical magnetic correlation length is
taken to be $\xiM=c/H_*\kpeak(1)$.
To compute the present values, we assume turbulent inverse cascading
at constant magnetic helicity until the matter-radiation equality using 
$\Brms^{\rm eq}=\Brms^* \eta_{\rm eq}^{-1/3}$ and
$\xiM^{\rm eq}=\xiM^* \eta_{\rm eq}^{2/3}$.
The value of $\eta_{\rm eq}$ is obtained by using
$g_{\rm eq}^{1/3} a_{\rm eq} T_{\rm eq} = g_{\rm r}^{1/3} a_{\rm r} \Tr$,
implied by the adiabatic evolution of the Universe and
$a_{\rm eq}=\eta_{\rm eq}$, where we take $T_{\rm eq}=1$eV and $g_{\rm eq}=3.94$.
The results are listed in \Tab{Ttoday}, where we use the
superscripts `r' and `eq' to indicate comoving values
at reheating and matter--radiation equality, respectively. 

We emphasize here that, unlike the magnetic field, which can have much
larger length scales owing to inverse cascading \citep{PFL76}, this is
not the case for GWs.
This is because GWs are governed by the imprint from the time
when the stress was maximum.

\section{Conclusions}
\label{Conclusions}

The present work has demonstrated that helical inflationary magnetogenesis
modifies the nonhelical case in such a way that the electric and magnetic
power spectra become strongly peaked at a finite wavenumber, corresponding
typically to about a tenth of the horizon scale at $\eta=1$.
Such a distinct wavenumber does not exist in the nonhelical case.
Except for the scale-invariant scaling in Run~C at superhorizon scales,
this leads to extremely blue spectra of electric and magnetic fields.
Nevertheless, the total stress has still always a purely white noise
spectrum and therefore also the GW field has a white noise spectrum
below its peak value.
Furthermore, for runs with large values of $\beta$, the onset of the
drop toward larger frequencies is much sharper in runs with helicity
than without.
These aspects can have observational consequences.
In particular, there would be more power at small wavenumbers and frequencies.
On the other hand, for a certain magnetic energy, helical magnetogenesis
produces somewhat weaker GWs than nonhelical magnetogenesis.
However, as we have shown here, the appropriate scaling is not with
$\EEM$, but with $\EEEM$, and therefore this conclusion is reversed.
In fact, the fractional contribution of electric fields to the
stress is much weaker in the helical case than without.

When studying GW generation from the CME, it was anticipated that some
general features or behaviors would carry over to other magnetogenesis
scenarios.
In magnetogenesis from the CME, the GW energy was well described by a
relation $\EEGW=(\qM\EEM/k_{\rm c})^2$, where the efficiency $\qM$
depended on the value of the conductivity and it also depended on
which of the two possible regimes one is in.
The possibility of two different regimes seems to be a special property of
the CME that has not yet been encountered in other magnetogenesis scenarios.
Also the presence of a conservation law of total chirality in the CME
has no obvious counterpart in inflationary magnetogenesis, where magnetic
helicity conservation is not obeyed during magnetogenesis in step~I.

On the other hand, both the CME and helical inflationary magnetogenesis
can produce circularly polarized GWs.
However, the CME operates only on very small length scales that are in
practice much smaller than what is shown in \Fig{ppol_comp}, where an
unphysically large chiral chemical potential was applied, just to see
what GW strengths would then be possible.
This naturally raises the question whether some combination of CME and
inflationary magnetogenesis could produce either stronger or larger
scale magnetic fields.
A problem lies in the fact that the CME requires electric conductivity.
It could therefore only be an effect that operates {\em after}
inflationary magnetogenesis and during the radiation-dominated era.
It could then enhance the magnetic field, but the resulting additional
magnetic field would then only be of short length scales.
Nevertheless, the preceding inflationary stage could lead to somewhat
stronger fields and could thereby also produce stronger GWs.
Another interesting effect could be the intermediate production of an
imbalance of fermions from the magnetic field produced by inflationary
magnetogenesis.
This aspect has recently been explored by \cite{Schober+2020}, who
showed that this effect is indeed only an intermediate one, because
at late times, the chiral imbalance always gets converted back into
magnetic fields.

When comparing a plot of $\EEGW$ versus $\EEM$ from inflationary
magnetogenesis, the work of BS has shown that a scaling of the
form $\EEGW\propto\EEM^2$ was obtained.
Our new results for helical inflationary magnetogenesis
explicitly confirm a $1/k_{\rm c}$ dependence, but here with
$\EEGW=(\qEM\EEEM/k_{\rm c})^2$, where $\qEM$ shows only a very
weak dependence on $\beta$.
Here, $k_{\rm c}=\kpeak(1)$ has been used (as in BS),
and $\qEM=1.1$--$1.6$ has been found as a fit parameter.
Note, however, that the formula for $\EEGW$ in terms of $\EEEM$ is
entirely empirical.
It would be important to produce some more robust analytic justification
or refinements to this expectation.

\begin{table*}\caption{
Model parameters for different values of $\Tr$.
}
\begin{tabular}{ccccccccccc} 
$\Tr$&$\alpha$&$\Iota$&$\EEEM$& $H_{\rm f}$ [GeV] &
$N_{\rm r}$&$N$&$\beta$&$g_{\rm r}$&$\EM(\eta_{\rm ini},k)$\\
\hline
$ 10\GeV$ &2&1&0.07& $2.3\times 10^{-11}$&8.1&31.1&7.7&86&$\propto k^3$\\
$  8\GeV$ &2&1&0.01 & $2.8 \times 10^{-11}$&8.6&31.1&7.3&86&$\propto k^3$\\
$120\MeV$ &2&1&0.01& $1.2\times 10^{-3}$&26.5&35.5&2.7&20&$\propto k^3$\\
$150\MeV$ &2&1&0.006& $2.7\times 10^{-4}$&24.5&35.1&2.9&61.75&$\propto k^3$\\
$460\GeV$ &-3&2.5&0.01& $1.7\times 10^{-8}$&7.3&32.9&3&106.75&$\propto k^{-1}$\\
 $3 \times 10^5$ GeV&1&1&0.01 & $ 10^{14}$&32.1&53.4&1.7&106.75&$\propto k^5$\\
\label{Tbeta2}
\end{tabular}
\end{table*}

Of observational interest may also be the profile and slope
with which ${\cal P}_{\rm GW}(k)$ increases at low $k$.
Interestingly, the fractional polarization continues to be nearly 100\%
for wavenumbers several times larger than the peak at $2\kpeak(1)$,
but shows a decline for smaller $k$.

\begin{acknowledgements}
We thank Tina Kahniashvili and Kandaswamy Subramanian for useful
discussions.
Nordita's support during the program on Gravitational Waves from the
Early Universe in Stockholm in 2019 is gratefully acknowledged.
This work was support through grants from the Swedish Research Council
(Vetenskapsradet, 2019-04234).
We acknowledge the allocation of computing resources provided by the
Swedish National Allocations Committee at the Center for Parallel
Computers at the Royal Institute of Technology in Stockholm and
Lindk\"oping.

\end{acknowledgements}

\vspace{2mm}\noindent
{\large\em Software and Data Availability.} The source code used for
the simulations of this study, the {\sc Pencil Code} \citep{JOSS},
is freely available on \url{https://github.com/pencil-code/}.
The DOI of the code is https://doi.org/10.5281/zenodo.2315093 {\tt v2018.12.16}
\citep{axel_brandenburg_2018_2315093}.
The simulation setup and the corresponding data are freely available on
\dataset[doi:10.5281/zenodo.5137202]{https://doi.org/10.5281/zenodo.5137202}; see also
\url{https://www.nordita.org/~brandenb/projects/HelicalMagnetoGenesisGW/}
for easier access of the same material as on the Zenodo site.

\appendix

\section{Relation between $\beta$ and the reheating temperature}
\label{RelationReheatingTemperature}

We discussed in \Sec{MagnetogenesisModel} various combinations of model
parameters $\beta$ and $\Iota$ for a chosen value of $\Tr$.
For the nonhelical case with $\Iota=0$, details were already
given in Appendix~A of BS.
The expression corresponding to Equation~(A1) of BS is obtained as follows.

Details of the helical magnetogenesis model are explained in SSS.
The expressions below their Equations~(23) and (29) represent the
solution for the scaled vector potential $\mathcal{A}_{h}$ during
inflation and the matter-dominated era, respectively, and are given by
\begin{align}
\mathcal{A}_{1h}(\eta) &= \frac{e^{-h \pi \aa/2}}{\sqrt{2 k}} W_{i \aa h, \aa +\frac{1}{2}}(2 i k \eta),\\
\mathcal{A}_{2h}(\zeta) &= d_1 M_{2 i \beta h,-(2 \beta +\frac{1}{2})}(2 i k \zeta)+ d_2 M_{2 i \beta h,
2 \beta +\frac{1}{2}}(2 i k \zeta).
\label{fullsol}
\end{align}
Here $h=\pm 1$, $\zeta$ is a time variable during the matter-dominated
era defined in SSS as $\zeta\equiv\eta-3\eta_{\rm f}$, where
$\eta_{\rm f}$ is the value of conformal time at the end of inflation,
and $W$ and $M$ represent the Whittaker functions of the first and
second kind.
The coefficients $d_1$ and $d_2$ are obtained by the matching
$A_{h}\equiv\mathcal{A}_{h}/f$ and its derivatives at the end of
inflation.
In SSS, only the $\mathcal{A}_{h}$ in the superhorizon
limit during the matter-dominated era was considered.
Since this solution does not incorporate the extra growth of
the modes when they start entering the horizon (as evident from
\Fig{rspec34P_P512b73b}), we consider the full solution given in
\Eq{fullsol} in the present paper.
By considering the full solution, we obtain $d_1$ and $d_2$
and, further using Equation~(29) in Equations~(17) and (18) of
SSS, we obtain the magnetic and electric energy densities
during the matter-dominates era.
Demanding that the total EM energy be smaller than the background energy
density at the end of inflation, we calculate the value of the Hubble
parameter during inflation, $H_{\rm f}$, for given values of $\Tr$,
$\alpha$, and $\EEEM$.
Further, using these values, we estimate the value of $\beta\equiv
2 N/N_r$, where $N$ and $N_r$ are the number of $e$-folds during the
post-inflationary matter-dominated era and during inflation, respectively.
We provide these values in \Tab{Tbeta2} along with the initial magnetic
field spectrum in the superhorizon limit during matter-dominated era
and the value of the relativistic degrees of freedom at the beginning
of the radiation-dominated era, $g_{\rm r}(\eta_*)$.

\bibliography{ref}{}

\begin{thebibliography}{}
\expandafter\ifx\csname natexlab\endcsname\relax\def\natexlab#1{#1}\fi
\providecommand{\url}[1]{\href{#1}{#1}}
\providecommand{\dodoi}[1]{doi:~\href{http://doi.org/#1}{\nolinkurl{#1}}}
\providecommand{\doeprint}[1]{\href{http://ascl.net/#1}{\nolinkurl{http://ascl.net/#1}}}
\providecommand{\doarXiv}[1]{\href{https://arxiv.org/abs/#1}{\nolinkurl{https://arxiv.org/abs/#1}}}

\bibitem[{Adshead {et~al.}(2016)Adshead, Giblin, Scully, \&
  Sfakianakis}]{adshead2016}
Adshead, P., Giblin, J.~T., Scully, T.~R., \& Sfakianakis, E.~I. 2016, JCAP,
  2016, 039, \dodoi{10.1088/1475-7516/2016/10/039}

\bibitem[{Adshead {et~al.}(2018)Adshead, Giblin, \& Weiner}]{adshead2018}
Adshead, P., Giblin, J.~T., \& Weiner, Z.~J. 2018, PhRvD, 98, 043525,
  \dodoi{10.1103/PhysRevD.98.043525}

\bibitem[{{Amaro-Seoane} {et~al.}(2017){Amaro-Seoane}, {Audley}, {Babak},
  {Baker}, {Barausse}, {Bender}, {Berti}, {Binetruy}, {Born}, {Bortoluzzi},
  {Camp}, {Caprini}, {Cardoso}, {Colpi}, {Conklin}, {Cornish}, {Cutler},
  {Danzmann}, {Dolesi}, {Ferraioli}, {Ferroni}, {Fitzsimons}, {Gair}, {Gesa
  Bote}, {Giardini}, {Gibert}, {Grimani}, {Halloin}, {Heinzel}, {Hertog},
  {Hewitson}, {Holley-Bockelmann}, {Hollington}, {Hueller}, {Inchauspe},
  {Jetzer}, {Karnesis}, {Killow}, {Klein}, {Klipstein}, {Korsakova}, {Larson},
  {Livas}, {Lloro}, {Man}, {Mance}, {Martino}, {Mateos}, {McKenzie},
  {McWilliams}, {Miller}, {Mueller}, {Nardini}, {Nelemans}, {Nofrarias},
  {Petiteau}, {Pivato}, {Plagnol}, {Porter}, {Reiche}, {Robertson},
  {Robertson}, {Rossi}, {Russano}, {Schutz}, {Sesana}, {Shoemaker}, {Slutsky},
  {Sopuerta}, {Sumner}, {Tamanini}, {Thorpe}, {Troebs}, {Vallisneri},
  {Vecchio}, {Vetrugno}, {Vitale}, {Volonteri}, {Wanner}, {Ward}, {Wass},
  {Weber}, {Ziemer}, \& {Zweifel}}]{2017arXiv170200786A}
{Amaro-Seoane}, P., {Audley}, H., {Babak}, S., {et~al.} 2017, arXiv e-prints,
  arXiv:1702.00786.
\newblock \doarXiv{1702.00786}

\bibitem[{Anand {et~al.}(2019)Anand, Bhatt, \& Pandey}]{anand2018}
Anand, S., Bhatt, J.~R., \& Pandey, A.~K. 2019, EPJC, 79, 119,
  \dodoi{10.1140/epjc/s10052-019-6619-5}

\bibitem[{{Anber} \& {Sorbo}(2006)}]{Anber+Sorbo06}
{Anber}, M.~M., \& {Sorbo}, L. 2006, \jcap, 2006, 018,
  \dodoi{10.1088/1475-7516/2006/10/018}

\bibitem[{{Arzoumanian} {et~al.}(2020){Arzoumanian}, {Baker}, {Blumer},
  {B{\'e}csy}, {Brazier}, {Brook}, {Burke-Spolaor}, {Chatterjee}, {Chen},
  {Cordes}, {Cornish}, {Crawford}, {Cromartie}, {Decesar}, {Demorest}, {Dolch},
  {Ellis}, {Ferrara}, {Fiore}, {Fonseca}, {Garver-Daniels}, {Gentile}, {Good},
  {Hazboun}, {Holgado}, {Islo}, {Jennings}, {Jones}, {Kaiser}, {Kaplan},
  {Kelley}, {Key}, {Laal}, {Lam}, {Lazio}, {Lorimer}, {Luo}, {Lynch},
  {Madison}, {McLaughlin}, {Mingarelli}, {Ng}, {Nice}, {Pennucci}, {Pol},
  {Ransom}, {Ray}, {Shapiro-Albert}, {Siemens}, {Simon}, {Spiewak}, {Stairs},
  {Stinebring}, {Stovall}, {Sun}, {Swiggum}, {Taylor}, {Turner}, {Vallisneri},
  {Vigeland}, {Witt}, \& {Nanograv Collaboration}}]{NANOGrav2020}
{Arzoumanian}, Z., {Baker}, P.~T., {Blumer}, H., {et~al.} 2020, \apjl, 905,
  L34, \dodoi{10.3847/2041-8213/abd401}

\bibitem[{Banerjee \& Jedamzik(2004)}]{banerjee2004}
Banerjee, R., \& Jedamzik, K. 2004, PhRvD, 70, 123003,
  \dodoi{10.1103/PhysRevD.70.123003}

\bibitem[{{Barnaby} {et~al.}(2011){Barnaby}, {Namba}, \& {Peloso}}]{Barnaby+11}
{Barnaby}, N., {Namba}, R., \& {Peloso}, M. 2011, \jcap, 2011, 009,
  \dodoi{10.1088/1475-7516/2011/04/009}

\bibitem[{{Biskamp} \& {M{\"u}ller}(1999)}]{BM99}
{Biskamp}, D., \& {M{\"u}ller}, W.-C. 1999, \prl, 83, 2195,
  \dodoi{10.1103/PhysRevLett.83.2195}

\bibitem[{{Boyarsky} {et~al.}(2012){Boyarsky}, {Fr{\"o}hlich}, \&
  {Ruchayskiy}}]{BFR12}
{Boyarsky}, A., {Fr{\"o}hlich}, J., \& {Ruchayskiy}, O. 2012, \prl, 108,
  031301, \dodoi{10.1103/PhysRevLett.108.031301}

\bibitem[{{Boyarsky} {et~al.}(2015){Boyarsky}, {Fr{\"o}hlich}, \&
  {Ruchayskiy}}]{BFR15}
---. 2015, \prd, 92, 043004, \dodoi{10.1103/PhysRevD.92.043004}

\bibitem[{Brandenburg(2018)}]{axel_brandenburg_2018_2315093}
Brandenburg, A. 2018, Pencil Code, v2018.12.16,  Zenodo,
  \dodoi{10.5281/zenodo.2315093}

\bibitem[{{Brandenburg} \& {Boldyrev}(2020)}]{BB20}
{Brandenburg}, A., \& {Boldyrev}, S. 2020, \apj, 892, 80,
  \dodoi{10.3847/1538-4357/ab77bd}

\bibitem[{{Brandenburg} {et~al.}(2021{\natexlab{a}}){Brandenburg}, {Clarke},
  {He}, \& {Kahniashvili}}]{2021arXiv210212428B}
{Brandenburg}, A., {Clarke}, E., {He}, Y., \& {Kahniashvili}, T.
  2021{\natexlab{a}}, PhRvD, in press, arXiv:2102.12428.
\newblock \doarXiv{2102.12428}

\bibitem[{{Brandenburg} {et~al.}(1996){Brandenburg}, {Enqvist}, \&
  {Olesen}}]{BEO96}
{Brandenburg}, A., {Enqvist}, K., \& {Olesen}, P. 1996, \prd, 54, 1291,
  \dodoi{10.1103/PhysRevD.54.1291}

\bibitem[{{Brandenburg} {et~al.}(2021{\natexlab{b}}){Brandenburg},
  {Gogoberidze}, {Kahniashvili}, {Mandal}, {Roper Pol}, \&
  {Shenoy}}]{BGKMRPS21}
{Brandenburg}, A., {Gogoberidze}, G., {Kahniashvili}, T., {et~al.}
  2021{\natexlab{b}}, CQGra, 38, 145002.
\newblock \doarXiv{2103.01140}

\bibitem[{{Brandenburg} {et~al.}(2021{\natexlab{c}}){Brandenburg}, {He},
  {Kahniashvili}, {Rheinhardt}, \& {Schober}}]{BHKRS21}
{Brandenburg}, A., {He}, Y., {Kahniashvili}, T., {Rheinhardt}, M., \&
  {Schober}, J. 2021{\natexlab{c}}, \apj, 911, 110 (BHKRS),
  \dodoi{10.3847/1538-4357/abe4d7}

\bibitem[{{Brandenburg} \& {Kahniashvili}(2017)}]{BK17}
{Brandenburg}, A., \& {Kahniashvili}, T. 2017, PhRvL, 118, 055102,
  \dodoi{10.1103/PhysRevLett.118.055102}

\bibitem[{Brandenburg {et~al.}(2017)Brandenburg, Kahniashvili, Mandal, Pol,
  Tevzadze, \& Vachaspati}]{axel2017}
Brandenburg, A., Kahniashvili, T., Mandal, S., {et~al.} 2017, \prd, 96, 123528,
  \dodoi{10.1103/PhysRevD.96.123528}

\bibitem[{{Brandenburg} {et~al.}(2017){Brandenburg}, {Kahniashvili}, {Mandal},
  {Pol}, {Tevzadze}, \& {Vachaspati}}]{Bran+17}
{Brandenburg}, A., {Kahniashvili}, T., {Mandal}, S., {et~al.} 2017, \prd, 96,
  123528, \dodoi{10.1103/PhysRevD.96.123528}

\bibitem[{{Brandenburg} \& {Sharma}(2021)}]{Bran+Shar21}
{Brandenburg}, A., \& {Sharma}, R. 2021, ApJ, in press, arXiv:2106.03857 (BS).
\newblock \doarXiv{2106.03857}

\bibitem[{Campanelli(2009)}]{campanelli2008}
Campanelli, L. 2009, IJMPD, 18, 1395, \dodoi{10.1142/S0218271809015175}

\bibitem[{{Caprini} \& {Sorbo}(2014)}]{Caprini+Sorbo14}
{Caprini}, C., \& {Sorbo}, L. 2014, \jcap, 2014, 056,
  \dodoi{10.1088/1475-7516/2014/10/056}

\bibitem[{{Caprini} {et~al.}(2016){Caprini}, {Hindmarsh}, {Huber},
  {Konstandin}, {Kozaczuk}, {Nardini}, {No}, {Petiteau}, {Schwaller},
  {Servant}, \& {Weir}}]{Caprini+16}
{Caprini}, C., {Hindmarsh}, M., {Huber}, S., {et~al.} 2016, \jcap, 2016, 001,
  \dodoi{10.1088/1475-7516/2016/04/001}

\bibitem[{Christensson {et~al.}(2001)Christensson, Hindmarsh, \&
  Brandenburg}]{axel2001}
Christensson, M., Hindmarsh, M., \& Brandenburg, A. 2001, PhRvE, 64, 056405,
  \dodoi{10.1103/PhysRevE.64.056405}

\bibitem[{Cornwall(1997)}]{cornwall1997}
Cornwall, J.~M. 1997, PhRvD, 56, 6146, \dodoi{10.1103/PhysRevD.56.6146}

\bibitem[{{Demozzi} {et~al.}(2009){Demozzi}, {Mukhanov}, \&
  {Rubinstein}}]{mukhanov2009}
{Demozzi}, V., {Mukhanov}, V., \& {Rubinstein}, H. 2009, JCAP, 8, 025,
  \dodoi{10.1088/1475-7516/2009/08/025}

\bibitem[{{Detweiler}(1979)}]{Detweiler79}
{Detweiler}, S. 1979, \apj, 234, 1100, \dodoi{10.1086/157593}

\bibitem[{{Domcke} {et~al.}(2020){Domcke}, {Ema}, \& {Mukaida}}]{Domcke+20}
{Domcke}, V., {Ema}, Y., \& {Mukaida}, K. 2020, JHEP, 2020, 55,
  \dodoi{10.1007/JHEP02(2020)055}

\bibitem[{Domcke \& Mukaida(2018)}]{domcke2018}
Domcke, V., \& Mukaida, K. 2018, JCAP, 2018, 020,
  \dodoi{10.1088/1475-7516/2018/11/020}

\bibitem[{Durrer {et~al.}(2011)Durrer, Hollenstein, \& Jain}]{durrer2011}
Durrer, R., Hollenstein, L., \& Jain, R.~K. 2011, JCAP, 2011, 037,
  \dodoi{10.1088/1475-7516/2011/03/037}

\bibitem[{{Ellis} {et~al.}(2020){Ellis}, {Fairbairn}, {Lewicki}, {Vaskonen}, \&
  {Wickens}}]{Ellis+20}
{Ellis}, J., {Fairbairn}, M., {Lewicki}, M., {Vaskonen}, V., \& {Wickens}, A.
  2020, \jcap, 2020, 032, \dodoi{10.1088/1475-7516/2020/10/032}

\bibitem[{{Ferreira} {et~al.}(2013){Ferreira}, {Jain}, \&
  {Sloth}}]{Ferreira+13}
{Ferreira}, R. J.~Z., {Jain}, R.~K., \& {Sloth}, M.~S. 2013, \jcap, 2013, 004,
  \dodoi{10.1088/1475-7516/2013/10/004}

\bibitem[{Fujita \& Durrer(2019)}]{fujita2019}
Fujita, T., \& Durrer, R. 2019, JCAP, 2019, 008,
  \dodoi{10.1088/1475-7516/2019/09/008}

\bibitem[{Fujita {et~al.}(2015)Fujita, Namba, Tada, Takeda, \&
  Tashiro}]{fujita2015}
Fujita, T., Namba, R., Tada, Y., Takeda, N., \& Tashiro, H. 2015, JCAP, 2015,
  054, \dodoi{10.1088/1475-7516/2015/05/054}

\bibitem[{Garretson {et~al.}(1992)Garretson, Field, \& Carroll}]{carroll1992}
Garretson, W.~D., Field, G.~B., \& Carroll, S.~M. 1992, PhRvD, 46, 5346,
  \dodoi{10.1103/PhysRevD.46.5346}

\bibitem[{{Gogoberidze} {et~al.}(2007){Gogoberidze}, {Kahniashvili}, \&
  {Kosowsky}}]{Gogo+07}
{Gogoberidze}, G., {Kahniashvili}, T., \& {Kosowsky}, A. 2007, \prd, 76,
  083002, \dodoi{10.1103/PhysRevD.76.083002}

\bibitem[{{Hatori}(1984)}]{Hat84}
{Hatori}, T. 1984, JPSJ, 53, 2539, \dodoi{10.1143/JPSJ.53.2539}

\bibitem[{{Hobbs} {et~al.}(2010){Hobbs}, {Archibald}, {Arzoumanian}, {Backer},
  {Bailes}, {Bhat}, {Burgay}, {Burke-Spolaor}, {Champion}, {Cognard}, {Coles},
  {Cordes}, {Demorest}, {Desvignes}, {Ferdman}, {Finn}, {Freire}, {Gonzalez},
  {Hessels}, {Hotan}, {Janssen}, {Jenet}, {Jessner}, {Jordan}, {Kaspi},
  {Kramer}, {Kondratiev}, {Lazio}, {Lazaridis}, {Lee}, {Levin}, {Lommen},
  {Lorimer}, {Lynch}, {Lyne}, {Manchester}, {McLaughlin}, {Nice}, {Oslowski},
  {Pilia}, {Possenti}, {Purver}, {Ransom}, {Reynolds}, {Sanidas}, {Sarkissian},
  {Sesana}, {Shannon}, {Siemens}, {Stairs}, {Stappers}, {Stinebring},
  {Theureau}, {van Haasteren}, {van Straten}, {Verbiest}, {Yardley}, \&
  {You}}]{Hobbs+10}
{Hobbs}, G., {Archibald}, A., {Arzoumanian}, Z., {et~al.} 2010, CQGra, 27,
  084013, \dodoi{10.1088/0264-9381/27/8/084013}

\bibitem[{Joyce \& Shaposhnikov(1997)}]{joyce1997}
Joyce, M., \& Shaposhnikov, M. 1997, PhRvL, 79, 1193,
  \dodoi{10.1103/PhysRevLett.79.1193}

\bibitem[{{Kahniashvili} {et~al.}(2021){Kahniashvili}, {Brandenburg},
  {Gogoberidze}, {Mandal}, \& {Pol}}]{2021PhRvR...3a3193K}
{Kahniashvili}, T., {Brandenburg}, A., {Gogoberidze}, G., {Mandal}, S., \&
  {Pol}, A.~R. 2021, PhRvR, 3, 013193, \dodoi{10.1103/PhysRevResearch.3.013193}

\bibitem[{Kahniashvili {et~al.}(2016)Kahniashvili, Brandenburg, \&
  Tevzadze}]{kahniashvili2016}
Kahniashvili, T., Brandenburg, A., \& Tevzadze, A.~G. 2016, PhysS, 91, 104008,
  \dodoi{10.1088/0031-8949/91/10/104008}

\bibitem[{{Kahniashvili} {et~al.}(2005){Kahniashvili}, {Gogoberidze}, \&
  {Ratra}}]{2005PhRvL..95o1301K}
{Kahniashvili}, T., {Gogoberidze}, G., \& {Ratra}, B. 2005, \prl, 95, 151301,
  \dodoi{10.1103/PhysRevLett.95.151301}

\bibitem[{{Kobayashi} \& {Afshordi}(2014)}]{Kobayashi+Afshordi14}
{Kobayashi}, T., \& {Afshordi}, N. 2014, JHEP, 2014, 166,
  \dodoi{10.1007/JHEP10(2014)166}

\bibitem[{{Kobayashi} \& {Sloth}(2019)}]{Kobayashi+Sloth19}
{Kobayashi}, T., \& {Sloth}, M.~S. 2019, \prd, 100, 023524,
  \dodoi{10.1103/PhysRevD.100.023524}

\bibitem[{{Maggiore}(2000)}]{Mag00}
{Maggiore}, M. 2000, \physrep, 331, 283, \dodoi{10.1016/S0370-1573(99)00102-7}

\bibitem[{{Mukhanov} {et~al.}(1992){Mukhanov}, {Feldman}, \&
  {Brandenberger}}]{Mukhanov+92}
{Mukhanov}, V.~F., {Feldman}, H.~A., \& {Brandenberger}, R.~H. 1992, \physrep,
  215, 203, \dodoi{10.1016/0370-1573(92)90044-Z}

\bibitem[{{Okano} \& {Fujita}(2021)}]{OF21}
{Okano}, S., \& {Fujita}, T. 2021, \jcap, 2021, 026,
  \dodoi{10.1088/1475-7516/2021/03/026}

\bibitem[{{Pencil Code Collaboration} {et~al.}(2021){Pencil Code
  Collaboration}, {Brandenburg}, {Johansen}, {Bourdin}, {Dobler}, {Lyra},
  {Rheinhardt}, {Bingert}, {Haugen}, {Mee}, {Gent}, {Babkovskaia}, {Yang},
  {Heinemann}, {Dintrans}, {Mitra}, {Candelaresi}, {Warnecke},
  {K{\"a}pyl{\"a}}, {Schreiber}, {Chatterjee}, {K{\"a}pyl{\"a}}, {Li},
  {Kr{\"u}ger}, {Aarnes}, {Sarson}, {Oishi}, {Schober}, {Plasson}, {Sandin},
  {Karchniwy}, {Rodrigues}, {Hubbard}, {Guerrero}, {Snodin}, {Losada},
  {Pekkil{\"a}}, \& {Qian}}]{JOSS}
{Pencil Code Collaboration}, {Brandenburg}, A., {Johansen}, A., {et~al.} 2021,
  JOSS, 6, 2807, \dodoi{10.21105/joss.02807}

\bibitem[{{Pouquet} {et~al.}(1976){Pouquet}, {Frisch}, \& {Leorat}}]{PFL76}
{Pouquet}, A., {Frisch}, U., \& {Leorat}, J. 1976, JFM, 77, 321,
  \dodoi{10.1017/S0022112076002140}

\bibitem[{{Ratra}(1992)}]{1992ApJ...391L...1R}
{Ratra}, B. 1992, \apjl, 391, L1, \dodoi{10.1086/186384}

\bibitem[{{Roper Pol} {et~al.}(2020{\natexlab{a}}){Roper Pol}, {Brandenburg},
  {Kahniashvili}, {Kosowsky}, \& {Mandal}}]{RoperPol+20b}
{Roper Pol}, A., {Brandenburg}, A., {Kahniashvili}, T., {Kosowsky}, A., \&
  {Mandal}, S. 2020{\natexlab{a}}, GApFD, 114, 130,
  \dodoi{10.1080/03091929.2019.1653460}

\bibitem[{{Roper Pol} {et~al.}(2021){Roper Pol}, {Mandal}, {Brandenburg}, \&
  {Kahniashvili}}]{RoperPol+21}
{Roper Pol}, A., {Mandal}, S., {Brandenburg}, A., \& {Kahniashvili}, T. 2021,
  JCAP, submitted, arXiv:2107.05356.
\newblock \doarXiv{2107.05356}

\bibitem[{{Roper Pol} {et~al.}(2020{\natexlab{b}}){Roper Pol}, {Mandal},
  {Brandenburg}, {Kahniashvili}, \& {Kosowsky}}]{RoperPol+20}
{Roper Pol}, A., {Mandal}, S., {Brandenburg}, A., {Kahniashvili}, T., \&
  {Kosowsky}, A. 2020{\natexlab{b}}, \prd, 102, 083512,
  \dodoi{10.1103/PhysRevD.102.083512}

\bibitem[{{Schober} {et~al.}(2020){Schober}, {Fujita}, \&
  {Durrer}}]{Schober+2020}
{Schober}, J., {Fujita}, T., \& {Durrer}, R. 2020, \prd, 101, 103028,
  \dodoi{10.1103/PhysRevD.101.103028}

\bibitem[{{Sharma} {et~al.}(2017){Sharma}, {Jagannathan}, {Seshadri}, \&
  {Subramanian}}]{Sharma+17}
{Sharma}, R., {Jagannathan}, S., {Seshadri}, T.~R., \& {Subramanian}, K. 2017,
  \prd, 96, 083511, \dodoi{10.1103/PhysRevD.96.083511}

\bibitem[{{Sharma} {et~al.}(2018){Sharma}, {Subramanian}, \&
  {Seshadri}}]{Sharma+18}
{Sharma}, R., {Subramanian}, K., \& {Seshadri}, T.~R. 2018, \prd, 97, 083503
  (SSS), \dodoi{10.1103/PhysRevD.97.083503}

\bibitem[{{Sharma} {et~al.}(2020){Sharma}, {Subramanian}, \&
  {Seshadri}}]{2020PhRvD.101j3526S}
---. 2020, \prd, 101, 103526, \dodoi{10.1103/PhysRevD.101.103526}

\bibitem[{{Taiji Scientific Collaboration} {et~al.}(2021){Taiji Scientific
  Collaboration}, {Luo}, {Wang}, {Bai}, {Bian}, {Cai}, {Cai}, {Cao}, {Chen},
  {Chen}, {Chen}, {Chen}, {Chen}, {Chen}, {Cong}, {Deng}, {Dong}, {Duan},
  {Fan}, {Fan}, {Fang}, {Fang}, {Feng}, {Feng}, {Feng}, {Gao}, {Gao}, {Guo},
  {He}, {He}, {Hou}, {Hu}, {Hu}, {Hu}, {Huang}, {Jia}, {Jiang}, {Jin}, {Jin},
  {Kang}, {Lei}, {Li}, {Li}, {Li}, {Li}, {Li}, {Li}, {Li}, {Li}, {Li}, {Li},
  {Li}, {Li}, {Li}, {Lin}, {Liu}, {Liu}, {Liu}, {Liu}, {Liu}, {Liu}, {Lu},
  {Luo}, {Ma}, {Ma}, {Ma}, {Ma}, {Man}, {Min}, {Niu}, {Peng}, {Peng}, {Qi},
  {Qiang}, {Qiao}, {Qu}, {Ruan}, {Sha}, {Shen}, {Shi}, {Shu}, {Su}, {Sui},
  {Sun}, {Tang}, {Tao}, {Tao}, {Tian}, {Wan}, {Wang}, {Wang}, {Wang}, {Wang},
  {Wang}, {Wang}, {Wang}, {Wang}, {Wang}, {Wei}, {Di Wu}, {Wu}, {Wu}, {Xi},
  {Xie}, {Xin}, {Xu}, {Xu}, {Xu}, {Xu}, {Xue}, {Xue}, {Yang}, {Yang}, {Yang},
  {Yang}, {Yang}, {Yang}, {Yin}, {Yu}, {Yu}, {Zhang}, {Zhang}, {Zhang},
  {Zhang}, {Zhang}, {Zhao}, {Zhao}, {Zhao}, {Zheng}, {Zhou}, {Zhu}, {Zou}, \&
  {Zou}}]{2021CmPhy...4...34T}
{Taiji Scientific Collaboration}, Wu, Y.-L., {Luo}, Z.-R., {Wang}, J.-Y.,
  {et~al.} 2021, CmPhy, 4, 34, \dodoi{10.1038/s42005-021-00529-z}

\bibitem[{{Turner} \& {Widrow}(1988)}]{1988PhRvD..37.2743T}
{Turner}, M.~S., \& {Widrow}, L.~M. 1988, \prd, 37, 2743,
  \dodoi{10.1103/PhysRevD.37.2743}

\bibitem[{Vachaspati(2001)}]{vachaspati2001}
Vachaspati, T. 2001, \prl, 87, 251302, \dodoi{10.1103/PhysRevLett.87.251302}

\bibitem[{{Vilenkin}(1980)}]{Vilenkin80}
{Vilenkin}, A. 1980, \prd, 22, 3080, \dodoi{10.1103/PhysRevD.22.3080}

\end{thebibliography}
\bibliographystyle{aasjournal}
\end{document}